\documentclass[journal,10pt,twocolumn,twoside]{IEEEtran}
\usepackage{amsmath,amsfonts,amssymb,amsbsy,bm,paralist,theorem,color}
\usepackage[T1]{fontenc}
\usepackage{amsmath}
\usepackage{algorithm,algpseudocode}
\usepackage{graphicx}
\graphicspath{ {./Figures/} }
\usepackage{multicol,lipsum}
\usepackage[cmintegrals]{newtxmath}
\usepackage{cuted}

\newcommand{\h} {\mathbf{h}}

\DeclareMathOperator*{\st}{s.t.}

\graphicspath{{fig/}}

\begin{document}
\title{Energy-Efficient IRS-Aided NOMA Beamforming for 6G Wireless Communications}
\author{Asim~Ihsan, Wen~Chen,~\IEEEmembership{Senior Member,~IEEE,} Muhammad~Asif, Wali~Ullah~Khan,~\IEEEmembership{Member,~IEEE}, and Jun~Li, ~\IEEEmembership{Senior Member,~IEEE}
	\thanks{(Corresponding author: Wen Chen.)}
	\thanks{Asim Ihsan and Wen Chen are with Department of Information and Communication Engineering, Shanghai Jiao Tong University, Shanghai 200240, China.	Email: \{ihsanasim;wenchen\}@sjtu.edu.cn.}
	\thanks{Muhammad Asif is with Guangdong Key Laboratory of Intelligent Information Processing, College of Electronics and Information Engineering, Shenzhen University, Shenzhen, Guangdong, China (email: masif@szu.edu.cn).}
	\thanks {Wali Ullah Khan is with the Interdisciplinary Centre for Security, Reliability and Trust (SnT), University of Luxembourg, 1855 Luxembourg City, Luxembourg (Emails: waliullah.khan@uni.lu).}
    \thanks {Jun Li is with the School of Electronic and Optical Engineering, Nanjing University of Science and Technology, Nanjing, 210094, CHINA (e-mail: jun.li@njust.edu.cn).}
}

\maketitle

\begin{abstract}
This manuscript presents an energy-efficient alternating optimization framework based on intelligent reflective surfaces (IRS) aided non-orthogonal multiple access beamforming (NOMA-BF) system for 6G wireless communications. Specifically, this work proposes a centralized IRS-enabled design for the NOMA-BF system to optimize the active beamforming and power allocation coefficient (PAC) of users at the transmitter in the first stage and passive beamforming at IRS in the 2nd stage to maximize the energy efficiency (EE) of the network. However, an increment in the number of supportable users with the NOMA-BF system will lead to NOMA user interference and inter-cluster interference (ICI). To mitigate the effect of ICI, first zero-forcing beamforming along with efficient user clustering algorithm is exploited and then NOMA user interference is tackled efficiently through a proposed iterative algorithm that computes PAC of NOMA user through simplified closed-form expression under the required system constraints. In the 2nd stage, the problem of passive beamforming is solved through a technique based on difference-of-convex (DC) programming and successive convex approximation (SCA). Simulation results demonstrate that the proposed alternating framework for energy-efficient IRS-assisted NOMA-BF system can achieve convergence within a few iterations and provide efficient performance in terms of EE of the system with low complexity.    
\end{abstract}
\vspace{-1em}
\section{INTRODUCTION AND MOTIVATIONS}
The wireless generations from 1G to 5G are designed through the assumption that the wireless environment cannot be controlled and modified. Therefore it can only be compensated through the design of efficient transmission/reception technologies. However, the enhancements at the endpoints of the wireless communications are not enough to meet the quality of services requirements of beyond 5G networks. Instead, 6G vision can be achieved by treating the wireless environment as an adaptable variable that is to be optimized. This can be possible through the use of Intelligent Reflective Surfaces (IRSs), which are known as enabling technology for 6G \cite{{R. Alghamdi}}. The 6G plan to serve billions of interconnected devices for applications like smart homes, cities and transportation through IRS would require advanced energy and spectral efficient resource allocation frameworks.

 IRSs have attracted considerable attention for their efficient control over electromagnetic waves. IRS contains an integrated electronic circuit with software that can control the propagation characteristics of the electromagnetic channel in a customizable way. It is a cost-effective technology for enhancing the energy and spectral efficiency of beyond 5G wireless communication which can be used in various new use cases. These include the utilization of IRS as a passive reflector, in which IRS operates without the connection of radio frequency (RF) chains and require very little power to control the tunable chips. The passive IRS can be deployed with scalable cost, lower energy consumption, and without sophisticated interference management \cite{Q. Wu 1}. Besides it, Active IRS is connected with radio frequency (RF) chains and can be used as active holographic MIMO which is the next step beyond massive MIMO because it has the potential to exploit the propagation characteristics of the electromagnetic channel with very low complexity in practical \cite{C. Huang,Z. Li}. Moreover, IRS has many favourable properties like the ease in integration into currently existing wireless networks and can be deployed on the buildings, roadside billboards, windows and indoor walls. These properties attracted researchers to investigate the benefits of IRS aided wireless communications, such as the transmit power minimization \cite{Q. Wu 2}, interference reduction \cite{D. Xu} and security enhancement \cite{X. Yu}.
 
 Non-orthogonal multiple access (NOMA) is a prominent multiple access candidate of the next-generation multiple access (NGMA) family for the 6G network \cite{Liu, A. Ihsan}. The integration of NOMA with IRS can achieve enhanced massive connectivity with higher spectral and energy efficiency. Moreover, IRS enabled NOMA networks are more adaptable than conventional NOMA networks because IRS can smartly adjust the direction of a channel vector of receivers, which assists in the implementation of NOMA \cite{Z. Ding}. In addition, IRS consist of multiple reflecting elements which can relax the strict constraint on the number of antennas at the transmitter and receiver in multiple antennas network \cite{Liu}. The above-mentioned merits of IRS-NOMA have drawn considerable research interest and several studies have been conducted. For instance, the authors in \cite{M. Fu} minimized the transmit power of the IRS-NOMA system by optimizing the beamforming vectors at BS and phase shift matrix at the IRS. In \cite{Z. Ding}, the performance of IRS assisted NOMA transmissions are described and investigated the influence of hardware impairments on their proposed design of IRS-NOMA system. Under the assumption of ideal and non-ideal IRS, the sum rate of the system is maximized in \cite{X. Mu} by the joint optimization of active beamforming at the source and passive beamforming at the IRS. The transmit power of IRS assisted multi-cluster MISO NOMA network is minimized in \cite{Y. Li}. The detailed comparison between IRS-NOMA with IRS-OMA is presented in \cite{B. Zheng}. Moreover, the interplay between IRS and NOMA is widely investigated for various technologies like unmanned aerial vehicle (UAV) \cite{X. Mu 2}, simultaneous wireless information and power transfer (SWIPT) \cite{Z. Li 2}, wireless powered communication networks (WPCNs) \cite{Q. Wu 3}, mobile edge computing (MEC) \cite{Q. Wang}, Backscatter Communications (BC) \cite{J. Zuo1}, and physical layer security (PLS) \cite{C. Gong}.  
 
To enable the internet of everything (IoE) applications through 6G wireless systems will result in a massive number of connections and huge data traffic. Therefore, the realization of IoE will require to gain high spectral efficiency \cite{M. Z. Chowdhury}. To further improve the sum capacity for future IoE applications, NOMA-BF can be exploited that can support multiple users in each cluster using a single BF vector through NOMA principles \cite{B. Kimy}. As IRS can smartly adjust the direction of a channel vector of receivers and it can also relax the strict constraint on the number of antennas at the transmitter and receiver in multiple antennas network. Therefore, it can be a prominent choice for the realization of IoE applications through NOMA-BF systems. The implementation of IRS aided NOMA-BF systems can be possible through centralized IRS-enabled design \cite{Liu Yuanwei,J. Zuo}.

Inspired and motivated by the above research contributions, the energy-efficient IRS-aided NOMA-BF system for 6G wireless communications is proposed in this paper. Our major contribution is summarized as follows, 
\begin{itemize}	
	\item \text{Energy-efficient centralized IRS NOMA-BF design:} The realization of IoE applications through 6G wireless systems for smart transportation, smart homes, and smart cities will result in the proliferation of wearable devices, mobile sensors, and electronic tablets. Therefore, this paper proposes a centralized IRS design along with the NOMA-BF system to increase the number of supportable users in an energy-efficient manner. The proposed centralized IRS design supports the users in each cell through multiple clusters and serves the multiple users in each cluster through one common beamformer through the NOMA principle. Although the number of supportable users increases with the NOMA-BF system, it results in NOMA user interference as well as in ICI. Therefore the proper interference management algorithms are essential to mitigate the degrading effects.
	
	\item \text{Interference management:} It is very vital to manage the interference in the IRS-enabled NOMA-BF system. This paper exploits zero-forcing beamforming along with an efficient user clustering algorithm to degrade the effects of ICI. The user clustering is based on maximum correlation and channel gain difference among users. The effect of the number of antennas at BS and passive elements at IRS on ICI is analyzed. Subsequently, the NOMA user interference is handled through a low-complexity energy-efficient power allocation algorithm that computes PAC of NOMA users in each cluster through simplified closed-form expression. 
	
	\item \text{Geometric channel model:} The analysis of the proposed NOMA-BF system for IRS-assisted communications is done under the consideration of a more practical geometric channel model. In the considered geometric model, the channel gains are modelled as rician fading, whose line-of-sight (LoS) component is presented by the array response of uniform linear array (ULA) of BS and IRS while the elements of the non-line-of-sight (NLoS) component are modelled as independently and identically distributed (i.i.d) complex Gaussian random variables with zero mean and unit variance.
	
	\item \text{Low complexity:} The proposed energy-efficient alternating optimization algorithm provides optimal EE performance for IRS aided  NOMA-BF systems in very low computational complexity. The detailed complexity analysis of the proposed algorithm is presented in section III. The performance of the proposed algorithm is analyzed through numerical Monte Carlo simulations and is compared with the benchmark algorithm. From obtained results, it is observed that the proposed algorithm can achieve higher EE performance with affordable computational complexity for practical implementations.
	 
\end{itemize}      


\section{System Model and Problem Formulation}
\begin{figure}
	\centering
	\includegraphics[width=80mm]{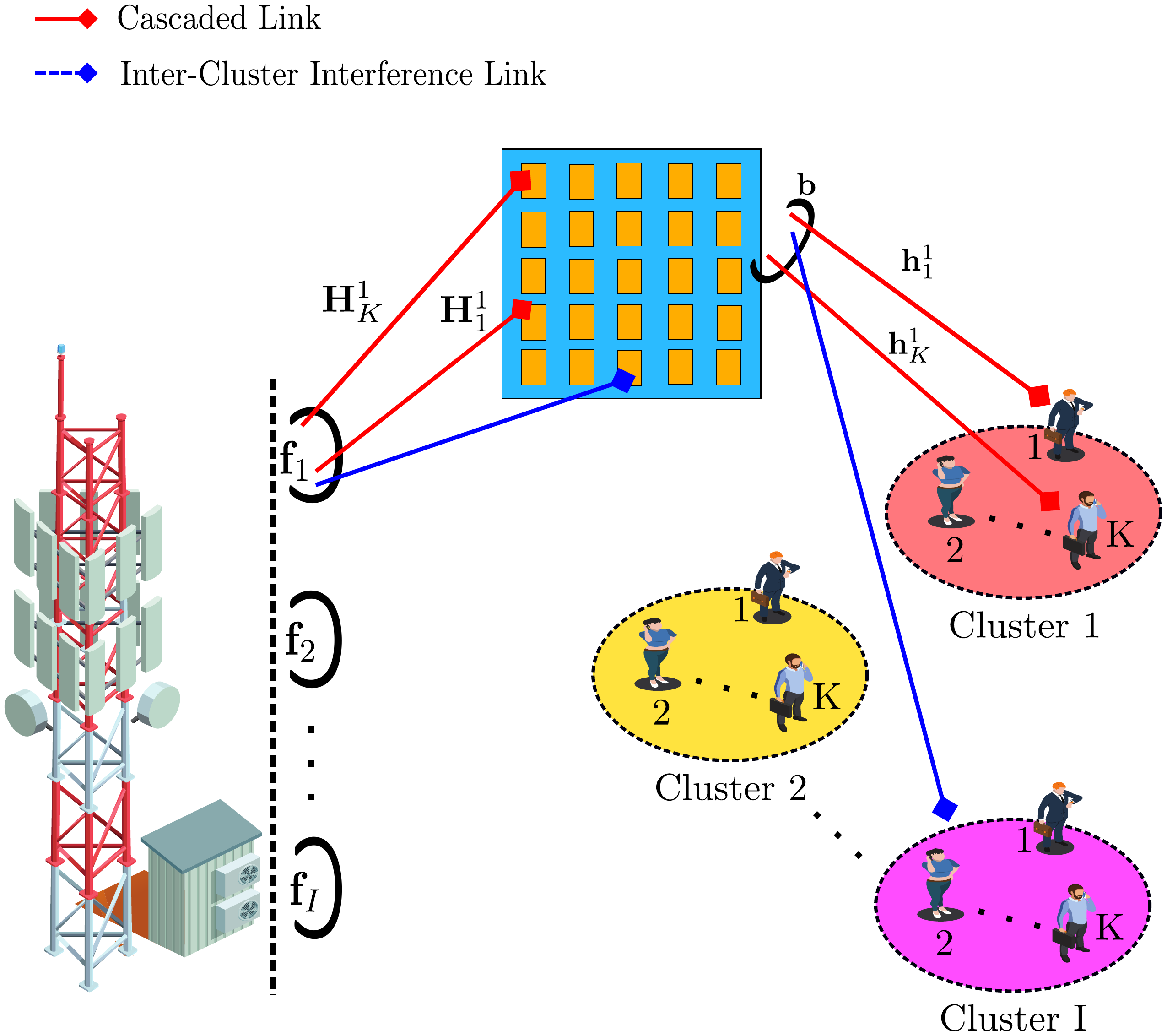}
	\caption{Illustration of Channel gains in IRS-Enabled NOMA-BF vector}
	\label{FIG:1}
\end{figure}
\begin{figure*}
	\centering
	\includegraphics[width=120mm]{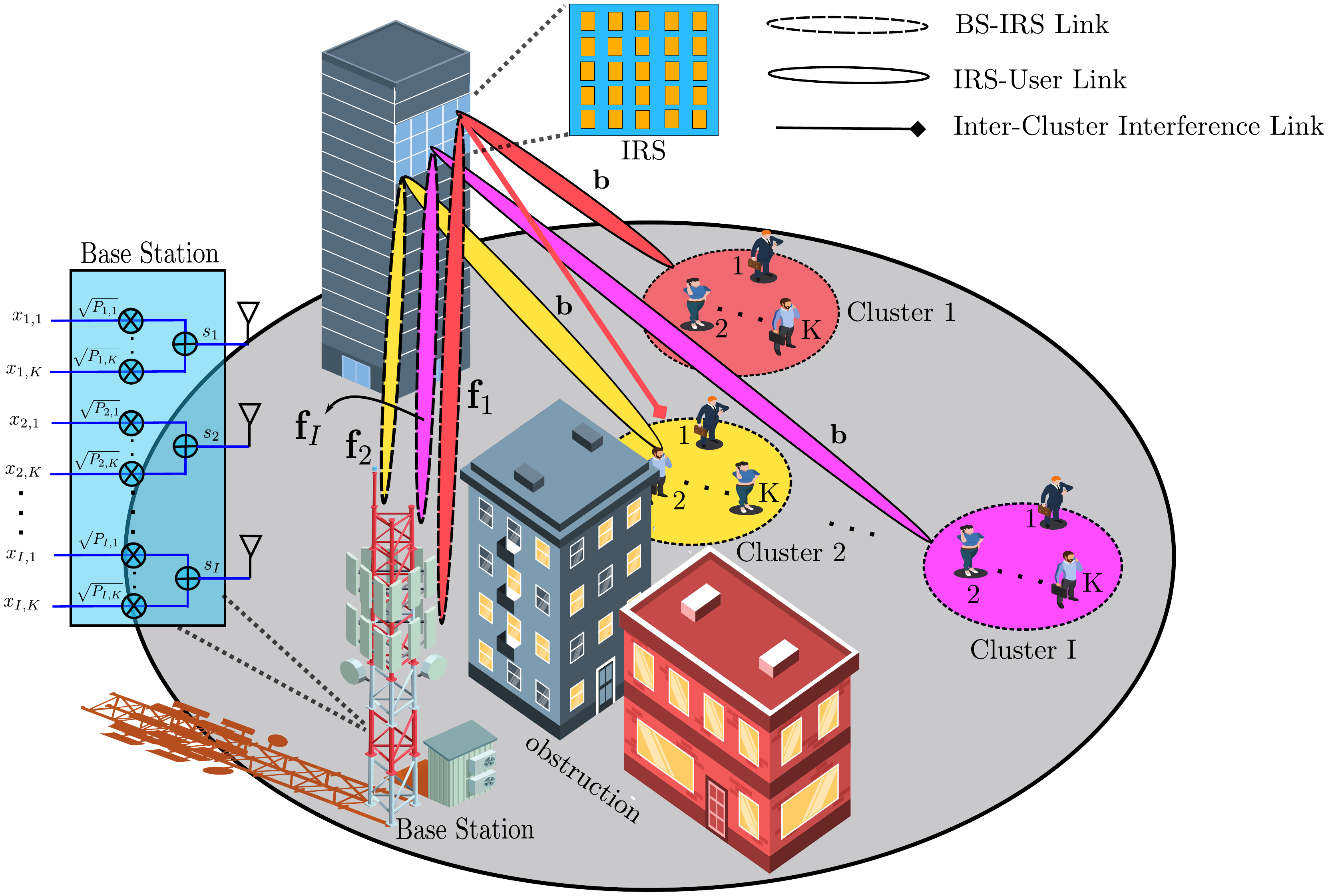}
	\caption{Illustration of System model}
	\label{FIG:2}
\end{figure*}

This work considers IRS-enabled downlink NOMA-BF system for beyond 5G wireless communications. The system is consist of one base station (BS) with $M$ uniform linear array (ULA) antennas and IRS with $N$ ULA reflecting elements, which is assisting downlink communications to $K$ dead-zone single-antenna receivers. The BS with $M$ antennas can transmit $\mathcal{I}$ BF vectors, where each BF vector can serve two or more receivers through NOMA principles. A group of receivers supported by single BF vector through NOMA (NOMA-BF) is defined as cluster. There are $I$ clusters in the system, which can support $KI$ receivers. The BS transmits  $\sum_{i=1}^I\mathbf{f}_is_i$, where $\mathbf{f}_i$ is the NOMA BF vector and ${s}_i$ is the superimposed information signal for NOMA receivers in the $i^{th}$ cluster. $s_i = \sum_{k=1}^{K} \sqrt{P_i\beta_{i,k}}x_{i,k}$, where $x_{i,k}$ is the information signal with $\mathbb{E}[|x_{i,k}|^2]=1$, $P_i$ is the transmit power of $i^{th}$ cluster and $\beta_{i,k}$ is the PAC of $k^{th}$ NOMA receiver $(r_{i,k})$. Let, $\mathbf{B} =diag(b_1,b_2,\ldots,b_n)$ denotes the passive beamforming matrix of IRS, where $b_n $ is the  reflection coefficient of the $n^{th}$ element of IRS. $ b_n= \beta_n e^{j\theta_n}$ where $\beta_n$ is the amplitude and $\theta_n$ is the phase shift of the reflection coefficient of IRS. In this system, the channel gains from the BS to IRS and from IRS to $r_{i,k}$ are modeled as rician fading,respectively, as follow
\begin{align}
	&	\overline{\mathbf{H}}^i_k = \sqrt{\frac{\delta}{1+\delta}}{{\mathbf{H}}^i_k}^{LoS}+\sqrt{\frac{1}{1+\delta}}{{\mathbf{H}}^i_k}^{NLoS}, \label{1}
\end{align}
and 
\begin{align}
	&\overline{\mathbf{h}}^i_{k} = \sqrt{\frac{\epsilon}{1+\epsilon}}{{\mathbf{h}}^i_{k}}^{LoS}+\sqrt{\frac{1}{1+\epsilon}}{{\mathbf{h}}^i_{k}}^{NLoS}, \label{2}
\end{align}
where ${{\mathbf{H}}^i_k}^{LoS} \in \mathbb{C}^{N\times M}$ and ${{\mathbf{h}}^i_{k}}^{LoS} \in \mathbb{C}^{N\times 1}$ are the line-of-sight (LoS) component of BS to IRS and from IRS to $r_{i,k}$ link, respectively. While ${{\mathbf{H}}^i_k}^{NLoS} \in \mathbb{C}^{N\times M}$ and ${{\mathbf{h}}^i_{k}}^{NLoS} \in \mathbb{C}^{N\times 1}$ are the non-line-of-sight (NLoS) component of BS to IRS and from IRS to $r_{i,k}$ link, respectively. $\delta$ is the Rician factor of BS to IRS link while $\epsilon$ is the Rician factor of IRS to $r_{i,k}$ link.
The LoS component ${{\mathbf{H}}^i_k}^{LoS}$ and ${{\mathbf{h}}^i_{k}}^{LoS}$ depends upon the array response of ULA and can be given as follows
\begin{align}
	&{{\mathbf{H}}^i_k}^{LoS} = A^H({\theta^{AoA}})A({\phi^{AoD}_{BS}}), \label{3}
\end{align}
and
\begin{align}
	&{{\mathbf{h}}^i_{k}}^{LoS} = A({\phi^{AoD}_{IRS}}), \label{4}
\end{align}
where $A({\theta^{AoA}})$ is the array response vector of ULA for angle of arrival (AoA) at the IRS, $A({\phi^{AoD}_{BS}})$ is the array response vector of ULA for angle of departure (AoD) at BS, and $A({\phi^{AoD}_{IRS}})$ is the array response of ULA for AoD at IRS. The array response of $\mathcal{E}$ elements ULA can be presented as 
\begin{align}
	&A(\Theta)=[1, e^{-j2\pi\frac{d}{\lambda}\sin{\Theta}},\ldots,e^{-j2\pi(\mathcal{E}-1)\frac{d}{\lambda}\sin{\Theta}}], \label{5}
\end{align}
where $\Theta$ can be AoA or AoD of the signal. $d$ and $\lambda$ are the element spacing and carrier wavelength. Each element of the NLoS components ${{\mathbf{H}}^i_k}^{NLoS} \in \mathbb{C}^{N\times M}$ and ${{\mathbf{h}}^i_{k}}^{NLoS} \in \mathbb{C}^{N\times 1}$ are independently and identically distributed (i.i.d) complex Gaussian random variables with zero mean and unit variance. Therefore, the channel gain of BS to IRS and IRS to $r_{i,k}$ link can be given as
\begin{align}
	&\mathbf{H}^i_k=\sqrt{L_0\Big(\frac{d_{B,R}}{d_0}\Big)^{-\alpha_{B,R}}}\overline{\mathbf{H}}^i_k, \label{6}
\end{align}
and 
\begin{align}
	&\mathbf{h}^i_{k}=\sqrt{L_0\Big(\frac{d^k_{R,U}}{d_0}\Big)^{-\alpha^k_{R,U}}}\overline{\mathbf{h}}^i_{k}, \label{7}
\end{align}
where $L_0$ denotes the path loss at the reference distance $d_0 = 1 (m)$, $d_{B,R}$ is the distance from the BS to the IRS, and $d^k_{R,U}$ is the distance from the IRS to the $r_{i,k}$. $\alpha_{B,R}$ and $\alpha^k_{R,U}$ are the path-loss exponents. The illustration for channel gains in IRS-enabled NOMA-BF vector is presented in Fig.~\ref{FIG:1} while the illustration of the whole system model is shown in Fig.~\ref{FIG:2}. Thus the received signal at receiver $r_{i,k}$ is given by 
\begin{align}
	y_{i,k}=&({\h^i_{k}}^H\mathbf{B}\mathbf{H}^i_k)\mathbf{f}_is_i+({\h^i_{k}}^H\mathbf{B}\mathbf{H}^i_k)\sum\limits_{{\substack{j=1 \\ j\neq i}}}^I\mathbf{f}_js_j+n_{i,k}, \label{8}
\end{align}
where $n_k$ is the AWGN with zero mean and variance ($\sigma^2$). Let, $\mathbf{b}=[b_1,b_2,\ldots, b_n]^T \in \mathbb{C}^{N\times 1}$ is the vector containing diagnol elements of matrix $\mathbf{B}$. Then Eq. \eqref{8}  can be written as
\begin{equation}
	y_{i,k}=(\mathbf{W}_{i,k}\mathbf{b}^H)\mathbf{f}_is_i+(\mathbf{W}_{i,k}\mathbf{b}^H)\sum\limits_{{\substack{j=1 \\ j\neq i}}}^I\mathbf{f}_js_j+n_{i,k}, \label{9}
\end{equation}
where $\mathbf{W}_{i,k}$ is the cascaded channel matrix from the BS to $r_{i,k}$ through the IRS and is presented as follow
	\begin{align}
		&\mathbf{W}_{i,k}=diag({\h^i_{k}}^H)\mathbf{H}^i_k. \label{10}
	\end{align}
As $s_i = \sum_{k=1}^{K} \sqrt{P_i \beta_{i,k}}x_{i,k}$, Inserting it in Eq. \eqref{9} results in  
\begin{align}
	y_{i,k}&=\underbrace{\mathbf{u}_{i,k}\mathbf{f}_i(\sqrt{P_i\beta_{i,k}}x_{i,k})}_\text{desired signal}+\underbrace{\mathbf{u}_{i,k}\mathbf{f}_i\sum\limits_{{\substack{l=1 \\ l\neq k}}}^K(\sqrt{P_i\beta_{i,l}}x_{i,l})}_\text{NOMA user interference}\nonumber\\
	&+\underbrace{\sum\limits_{{\substack{j=1 \\ j\neq i}}}^I\mathbf{u}_{i,k}\mathbf{f}_j(\sum_{k=1}^{K} \sqrt{P_j\beta_{j,k}}x_{j,k})}_\text{Inter-cluster Interference}+\underbrace{n_{i,k}}_\text{noise}, \label{11}
\end{align}
where 
\begin{align}
	&\mathbf{u}_{i,k}=\mathbf{W}_{i,k}\mathbf{b}^H. \label{12}
\end{align}
Without loss of generality, the channel gain of the receivers can be sorted as $||\mathbf{u}_{i,k+1}||^2 \geq||\mathbf{u}_{i,k}||^2$. Following the consider channel order, the received SINR of $r_{i,k}$ after SIC is as follow
  is given as
   \begin{align}
  	&	\gamma_{i,k}=\frac{P_{i} \beta_{i,k}| \mathbf{u}_{i,k}\mathbf{f}_i|^2} {P_i\sum\limits_{l=k+1}^{K}\beta_{i,l}|\mathbf{u}_{i,k}\mathbf{f}_i|^2+\Psi_{i}+\sigma^2}, \label{13}
  \end{align}
where $\Psi_{i}$ denotes ICI for $i^{th}$ cluster and is given as follow
\begin{align}
	&\Psi_{i}={\sum\limits_{{\substack{j=1 \\ j\neq i}}}^I|\mathbf{u}_{i,k}\mathbf{f}_j|^2( \sum_{k=1}^{K}P_j\beta_{j,k})}. \label{14}
\end{align}
 Then , the achievable rate of $r_{i,k}$ can be written as
 \begin{align}
 	&R_{i,k}=BW\log_2(1+\gamma_{i,k}). \label{15}
 \end{align}
Accordingly, the sum-rate of the each cluster is obtained as
\begin{align}
	&R_i=\sum\limits_{k=1}^K R_{i,k}. \label{16}
\end{align}
and its corresponding power consumption is 
\begin{align}
	&P_{i,T}={\|\mathbf{f}_i\|^2 P_i\sum\limits_{k=1}^K \beta_{i,k}+P_c}, \label{17}
\end{align}
where $P_c$ is the circuit power consumption. The objective is to maximize the energy efficiency (EE) of the IRS aided NOMA-BF system through the optimization of the active beamforming and PAC of users at the BS and reflection coefficients vector of IRS for all receivers. From Eq. \eqref{11}, it can be observed that sharing a BF vector through NOMA principles lead to ICI as well as NOMA user interference. Therefore, to mitigate all kinds of interferences and improve energy efficiency, this paper exploits zero-forcing beamforming and an efficient clustering algorithm for mitigation of ICI and power allocation algorithm to reduce NOMA user interference. Additionally, an energy-efficient passive beamforming algorithm for IRS is also presented that further enhance the EE of the proposed IRS aided NOMA-BF system. Under the QoS requirements of NOMA receivers, power constraint for efficient SIC implementation \cite{M. S. Ali}, transmit power constraint at BS and reflection coefficients constraint at IRS, the optimization problem for the proposed system can be expressed as follows
\begin{subequations}\label{Prob:EE_1}
	\begin{align}
		\mathop {\max }\limits_{\bm{\beta},\mathbf{f}_{i},\mathbf{b}}& EE=  \mathop {\max }\limits_{\bm{\beta},\mathbf{f}_{i},\mathbf{b}}  \sum\limits_{i=1}^I  \frac{R_i}{P_{i,T}}\\
		\st\ & \gamma_{i,k}\geq \gamma^{\min}_{i,k} \label{EE_1-C1},\forall k,i,\\
		& \|\mathbf{f}_i\|^2P_i\beta_{i,k}|\mathbf{u}_{i,k+1}|^2-\|\mathbf{f}_i\|^2\sum\limits_{l=k+1}^KP_i\beta_{i,l}|\mathbf{u}_{i,k+1}|^2 \geq P_{g}  \nonumber\\
		&  ,k= \{1,2 ,...,K-1\},\forall i,   \label{EE_1-C2}\\
		&\|\mathbf{f}_i\|^2 P_i\sum\limits_{k=1}^K \beta_{i,k} \leq P_{\max}, \label{EE_1-C3}\forall k,i,\\
		& |{b}_{n}|= 1,\label{EE_1-C4} \forall n, 
	\end{align}
\end{subequations}
where $EE$ is the energy efficiency and $\bm{\beta}=\{\beta_{i,1},\beta_{i,2},\ldots,\beta_{i,K}\}$ is the vector of PAC of NOMA receivers. Constraint \eqref{EE_1-C1} is the QoS requirement constraint of the receivers while constraint \eqref{EE_1-C2} is power constraint for efficient SIC implementation in NOMA-BF cluster and $P_g$ is the minimum power gap required  to  differentiate  between  the information  to  be decoded and the remaining non-decoded information. The constraint \eqref{EE_1-C3} limits the transmit power of BS to $P_{max}$. The constraint \eqref{EE_1-C4} is the reflection  coefficient constraint of the IRS. 

\section{The solution of formulated problem}

The EE maximization problem defined in Eq. \eqref{Prob:EE_1} is a non-concave function, which is coupled with optimization variables such as beamforming vectors, PAC of users and IRS reflection coefficients. Thus it is very difficult to tackle it directly. To cope with this challenging problem, this work proposes a novel alternating optimization iterative algorithm. The proposed algorithm deals with the optimization problem in two stages. In the first stage, it exploits zero-forcing beamforming and optimizes the PAC of users through the fix reflection coefficient of IRS while in the 2nd stage it computes the optimal reflection coefficients of IRS with optimal active beamformers and PAC of users obtained in stage 1.

\subsection{Stage 1: Optimal Beamforming and PAC of the users}
For the given reflection coefficients of IRS, the optimization problem in \eqref{Prob:EE_1} can be simplified to a subproblem of active beamforming design along with PAC of users as 

\begin{subequations}\label{Prob:EE_2}
	\begin{align}
		\mathop {\max }\limits_{\bm{\beta},\mathbf{f}_{i}} EE=
		&\mathop {\max }\limits_{\bm{\beta},\mathbf{f}_{i}} \sum\limits_{i=1}^I  \frac{R_i}{P_{i,T}}\\
		\st\ \quad & \eqref{EE_1-C1},\eqref{EE_1-C2}, and \eqref{EE_1-C3}.  
	\end{align}
\end{subequations}
The above optimization problem has a non-concave objective function with its non-convex constraint set. Therefore, it is difficult to solve this non-convex optimization problem. For its efficient solution, first, we explored zero-forcing (ZF) beamforming, which nulls the ICI at the NOMA-BF user, whose channel gain is used to generate the beamforming vector of the $I^{th}$ cluster. However, the other NOMA users in the same cluster do receive the ICI. The beamforming vectors of the clusters are generated through the channels of highest channel gain receivers as $\mathbf{U}_i=[\mathbf{u}_{1,K},\ldots, \mathbf{u}_{i-1,K},\mathbf{u}_{i+1,K}\ldots,\mathbf{u}_{I,K}]$. ZF beamforming will induce ZF constraint for the optimization problem defined in \eqref{Prob:EE_2} as follows
\begin{align}
	&{\mathbf{U}_i}^H\mathbf{f}_i = 0, \forall i\label{20}.
\end{align}
Now the optimization problem in Eq. \eqref{Prob:EE_2} can be reformulated by exploiting the problem with ZF constraint and $\|{\mathbf{f}}_i\|^2 =1$, then, it can be represented as follow
\begin{subequations}\label{Prob:EE_3}
	\begin{align}
		\mathop {\max }\limits_{\bm{\beta},\mathbf{f}_{i}} EE=
		&\mathop {\max }\limits_{\bm{\beta},\mathbf{f}_{i}} \sum\limits_{i=1}^I  \frac{R_i}{P_{i,T}} \label{14_a}\\
		\st\ \quad &\eqref{EE_1-C1},\eqref{EE_1-C2}, and \eqref{EE_1-C3} \\
		&{\mathbf{U}_i}^H{\mathbf{f}}_i= 0, \forall i \label{14_c},\\
		&\|{\mathbf{f}}_i\|^2 =1, \forall i\label{14_d}, 
	\end{align}
\end{subequations}
According to preposition 5.1 in \cite{Q. Shi}, the optimal solution of ${\mathbf{f}}_i$ for problem \eqref{Prob:EE_3} results in
\begin{align}
	&{\mathbf{f}}_i = \frac{\mathbf{Q}_i\mathbf{Q}_i^H\mathbf{u}_{i,K}}{\|\mathbf{Q}_i\mathbf{Q}_i^H\mathbf{u}_{i,K}\|^2} \label{Eq_22},
\end{align}
where $\mathbf{Q}_i$ is the orthogonal basis of the null space of $\mathbf{U}_i^H$. It can be observed that the objective function \eqref{14_a} increases with the increase in the values of the terms $|\mathbf{u}_{i,k}\mathbf{f}_i|^2$ and with the decrease in the term of ICI. Therefore, it should be noted that as beamforming vectors for clusters are generated through the channel of highest channel gain users, therefore the optimal value of ${\mathbf{f}}_i$ computed through Eq. \eqref{Eq_22} will maximize the $|\mathbf{u}_{i,k}\mathbf{f}_i|^2$ term for highest channel gain user and will also null its ICI term but for the other users in the cluster we still need to implement an efficient user clustering algorithm that should select those users in the cluster with the highest channel gain user that have maximum channel correlation with it under the maximum channel gain difference. The maximum channel correlation among users in the same cluster will result in maximizing $|\mathbf{u}_{i,k}\mathbf{f}_i|^2$ term and reducing the ICI term for all users while maintaining maximum channel gain difference among them will foster the practical implementation of NOMA-BF system. After obtaining the optimal active beamformers, and implementing the efficient user clustering as presented in algorithm 1, the optimization problem in \eqref{Prob:EE_3} can be simplified as follow,
\begin{subequations}\label{Prob:EE_4}
	\begin{align}
		&\mathop {\max }\limits_{\bm{\beta}} EE =
		\mathop {\max }\limits_{\bm{\beta}} \sum\limits_{i=1}^I  \frac{R_i}{P_{i,T}} \label{16_a}\\
		\st\ & \gamma_{i,k}\geq \gamma^i_{k,\min} \label{16-C1},\forall k,i,\\
		&P_i\beta_{i,k}|\mathbf{u}_{i,k+1}|^2-\sum\limits_{l=k+1}^KP_i\beta_{l,k}|\mathbf{u}_{i,k+1}|^2 \geq P_{g}  \forall k,i, \label{16-C2}\\
		& P_i\sum\limits_{k=1}^K \beta_{i,k} \leq P_{\max}, \label{16-C3}\forall k,i,
	\end{align}
\end{subequations}

	\begin{algorithm}
	\caption{User-clustering for IRS-enabled NOMA-BF system for mitigation of ICI}
	\label{pseudoPSO}
	\begin{algorithmic}[1]
		\State {}\textbf{Initialize:} Total number of users in a cell $\mathcal{V}$, number of users per cluster K, number of clusters $\mathcal{I}$, x=1, i=1 and threshold $\Delta$ where $(0 \geq \Delta \leq 1)$. 
		\State Generate channel gain of all users for IRS-enabled NOMA-BF system. i.e. $G = [\mathbf{u}_{1}, \mathbf{u}_{2}, \ldots, \mathbf{u}_{v}, \ldots,\mathbf{u}_{V}]$. Where $\mathbf{u}_{v}=\mathbf{W}_{v}\mathbf{b}_{v}^H$ and $\mathbf{W}_{v}=diag({\h_{v}}^H)\mathbf{H}_v$
		\State \textbf{Stage:1} Compute the channel gain difference of user with all other users in the cell under the pre-defined correlation threshold. The channel gain difference of all users with respect to others should be computed as follow,
		\While {$x \leq \mathcal{V}$}
		\For{y=x+1:$\mathcal{V}$}
		\State $C(x,y)= \frac{|\mathbf{u}_{x}^H\mathbf{u}_{y}|}{|\mathbf{u}_{x}^H||\mathbf{u}_{y}|}$
		\If{$C(x,y) > \Delta$}
		\State $D(x,y)= ||u_x||^2-||u_y||^2$
		\Else
		\State $D(x,y)=0$
		\EndIf
		\EndFor
		\State $x=x+1$
		\EndWhile
		\State \textbf{Output:} A matrix $\mathbf{D}$.
		\State \textbf{Stage:2} User grouping
		\While {$i \leq \mathcal{I}$}
		\State $Cluster_i$ = $max (\mathbf{D})$ selects the users which has highest channel gain difference among them under the predefined channel correlation threhold.
		\State Remove the selected users channel gain difference entries with all other users from $\mathbf{D}$.
		\State $i=i+1$
		\EndWhile
		\State \textbf{Output:} $\mathcal{I}$ Clusters.
	\end{algorithmic}
\end{algorithm}
 
The objective function of the above optimization problem has a non-linear fractional form, which is non-convex for the optimization variable $\beta_{i,k}$. Therefore, logarithmic approximation \cite{J. Papandriopoulos} is adopted to tackle this challenging problem. This successive convex approximation (SCA)  approximates the sum-rates expression in each iteration as follow
\begin{equation}
	\zeta\log_{2}(\gamma)+\Omega \leq \log_{2}(1+\gamma), \label{eq:24}
\end{equation}
where $ \zeta = \frac {\gamma_0}{1+\gamma_0} $ and $\Omega=\log_{2}(1+{\gamma_0})-\frac {\gamma_0}{1+\gamma_0}\log_{2}(\gamma_0) $. When $\gamma =\gamma_0$, the bound becomes tight. Through lower bound of inequality in Eq. (\ref{eq:24}), the sum-rate of IRS NOMA-BF system is presented as
\begin{equation}
	\overline{R}_i= \sum_{k=1}^{K}BW(\zeta_{i,k}\log_{2}({ \gamma_{i,k}})+\Omega_{i,k}), \label{eq:25}
\end{equation}
where
\begin{equation}
	\zeta_{i,k}= \frac {{\gamma_{i,k}}}{1+{\gamma_{i,k}}}, \label{eq:26}
\end{equation}
and
\begin{equation}
	\Omega_{i,k}=\log_{2}(1+{\gamma_{i,k}})-\frac {{ \gamma_{i,k}}}{1+{ \gamma_{i,k}}}\log_{2}({\gamma_{i,k}}). \label{eq:27}
\end{equation}
Based on the above logarithmic approximation, the optimization
problem in Eq. \eqref{Prob:EE_4}  can be rewritten as
\begin{subequations}\label{Prob:EE_5}
	\begin{align}
		\mathop {\max }\limits_{\bm{\beta}} EE =
		&\mathop {\max }\limits_{\bm{\beta}} \sum\limits_{i=1}^I  \frac{\overline{R}_i}{P_{i,T}} \label{21_a}\\
		\st\ \quad & \eqref{16-C1},\eqref{16-C2}, and \eqref{16-C3}.		
	\end{align}
\end{subequations}
Now, the above problem can be solved in an affordable complexity through Dinkelbach’s algorithm \cite{K. Shen}, which transforms the fractional objective function in Eq. \eqref{21_a} to
parametric form as
	\begin{align}
			\mathop {\max }\limits_{\bm{\beta}} EE =\max_{\bm{\beta}}\: &\sum\limits_{i=1}^I F(\rho_i)=\mathop  {\max }\limits_{\bm{\beta}} \sum\limits_{i=1}^I  {\overline{R}_i}-\rho_i P_{i,T},\label{29} 
	\end{align}
where, $\rho_i$ presents the maximum EE of $i^{th}$ cluster. The computation of roots of $F(\rho_i)$ is analogous to the solution of the objective function in Eq. \eqref{21_a} \cite{M. R. Zamani}. $F(\rho_i)$ is convex for $\rho_i$ because $F(\rho_i)$ is negative when $\rho_i$ approaches infinity, while $F(\rho_i)$ is positive when $\rho_i$ approaches minus infinity.  The convex problem in Eq. \eqref{29} is solved by employing the special case of Lagrangian relaxation, which is known as dual decomposition. The Lagrangian function of the optimization problem in \eqref{29}  presented as follow
\begin{align} 
	&\mathcal{L}(\bm{\beta},\alpha_i, \bm{\varphi_i},\bm{\Upsilon_i})= {\overline{R}_i}-\rho_i P_{i,T}+\alpha_i(P_{max}-P_i\sum\limits_{k=1}^K\beta_{i,k}) \nonumber\\
	&+\sum\limits_{k=1}^K\varphi_{i,k}(\gamma_{i,k}- \gamma^{\min}_{i,k})+\sum\limits_{k=1}^{K-1}\Upsilon_{i,k}(P_i\beta_{i,k}|\mathbf{u}_{i,k+1}|^2 \nonumber\\
	&-\sum\limits_{l=k+1}^KP_i\beta_{i,l}|\mathbf{u}_{i,k+1}|^2- P_{g}), \label{30}
\end{align} 
where $\alpha$, $\boldsymbol{\varphi_i}=\{\varphi_{i,1},\varphi_{i,2}, \cdots,\varphi_{i,K} \}$, and $\boldsymbol{\Upsilon_i}=\{\Upsilon_{i,1},\Upsilon_{i,2},\cdots,\Upsilon_{i,K} \}$ are the Lagrange multipliers. Constraints are KKT conditions for optimizing the power allocation of NOMA-BF users.
\\
\\
$\boldsymbol{Lemma 1:}$
The closed-form solution of optimal PAC of NOMA-BF user can be expressed as
\begin{align}
	\beta_{i,k}= \frac{{BW }\zeta_{i,k}}{\ln{2}((\rho_i+\alpha_i )P_i-\Gamma_{i,k})+\sum_{z=1}^{k-1}\Pi(\beta_{i,z})},  \label{eq:31}
\end{align}
where 
\begin{align}
	\Gamma_{i,k} = \varphi_{i,k}P_{i}|\mathbf{u}_{i,k}\mathbf{f}_i|^2-\Upsilon_{i,k}|\mathbf{u}_{i,k+1}|^2,  \label{eq:32}
\end{align}
and 
\begin{align}
	\Pi(\beta_{i,z}) = \Big(\frac{BW\zeta_{i,z}}{\beta_{i,z}}\Big)\gamma_{i,z}+\varphi_{i,z}P_{i}|\mathbf{u}_{i,z}\mathbf{f}_i|^2\gamma^{\min}_{i,z}+\Upsilon_{i,z}|\mathbf{u}_{i,z+1}|^2.  \label{eq:33}
\end{align}
$\boldsymbol{Proof:}$ Please, refer to Appendix A 

 Given the optimal PAC policy in Eq. (\ref{eq:31}), the primal problem's dual variables can be computed and updated iteratively by using the sub-gradient method.
\begin{align}
	\alpha_i(iter+1)=&\Big[\alpha_i(iter)-\omega_1(iter)\Big(P_{max}-P_i\sum\limits_{k=1}^K\beta_{i,k}\Big)\Big]^+, \label{34} \\
	\varphi_{i,k}(iter+1)=&\Big[\varphi_{i,k}(iter)-\omega_2(iter)\Big({P_{i} \beta_{i,k}| \mathbf{u}_{i,k}\mathbf{f}_i|^2}-\nonumber\\
	&\gamma^{\min}_{i,k}\big({P_i\sum\limits_{l=k+1}^{K}\beta_{i,l}|\mathbf{u}_{i,k}\mathbf{f}_i|^2+\Psi_{i}+\sigma^2}\big)\Big)\Big]^+, \label{35} \\
	\Upsilon_{i,k}(iter+1)=&\Big[\Upsilon_{i,k}(iter)-\omega_3(iter)\Big(P_i\beta_{i,k}|\mathbf{u}_{i,k+1}|^2- \nonumber \\
	&\sum\limits_{l=k+1}^KP_i\beta_{i,l}|\mathbf{u}_{i,k+1}|^2- P_{g}\Big)\Big]^+,  \label{36} 
\end{align}
where $iter$ is used for iteration index. $\omega_1$, $\omega_2$, and $\omega_3$ present positive step sizes. It is important to use the appropriate step sizes for the convergence to an optimal solution.
\subsection{Stage 2: Efficient Reflection Coefficients of IRS}
\begin{algorithm}
	\caption{Energy-Efficient Resource Allocation for IRS-enabled NOMA-BF system }
	\label{pseudoPSO}
	\begin{algorithmic}[1]
		\State \textbf{Stage 1: Optimal Beamforming and PAC of the Users}  
		\State \textbf{Initialization:} Initialize the PAC for each NOMA-BF user,  stepsizes, dual variables, maximum iterations $L_{max}$, iteration index $l = 1$ and maximum tolerance $\delta_{max}$.
		\For{$l \leq L_{max} $}
		\State Compute ${\overline{R}_i}(l)$ by using Eq. (\ref{eq:25})
		\State Compute $\rho_i(l)= \frac {\overline{R}_i(l)} {\|\mathbf{f}_i\|^2 P_i\sum\limits_{k=1}^K \beta_{i,k}+P_c}$
		\State Update dual variables $\alpha_i(l)$, $\bm{\varphi_i}(l)$, and $\bm{\Upsilon_i}(l)$ by using Eq. (\ref{34}), (\ref{35}) and (\ref{36}), respectively.
		\State Update the PAC vector $\boldsymbol{\beta}(l+1)$ of IRS NOMA-BF users by using equation (\ref{eq:31})
		\If{$|F^{(l)}_i(\rho)-F^{(l-1)}_i(\rho)| \leq \delta_{max}$}
		\State {\textbf{break}};
		\EndIf
		\EndFor\\
		\textbf{Output:} Optimal $\boldsymbol{\beta}^*=\{\beta_{i,1}^*,\beta_{i,2}^*,\cdots,\beta_{i,K}^*\}$
		\State \textbf{Stage 2: Optimal reflection coefficients algorithm (ORCA) for IRS NOMA-BF system}
		\State  \textbf{Problem Transformation:} Transform the IRS RC optimization problem defined in into tractable form through following steps.\\
		1) \textbf{DC programming}: Apply DC programming approch on the $\overline{R}^*_i$ using Eq. \eqref{eq:47} \\
		2)\textbf{Rank one constraint approximation}: Transform the non-convex rank one constraint in the form of difference of two convex functions using Eq \eqref{eq:49} and then apply SCA on it using using Eq. \eqref{eq:50}.\\
		3)\textbf{Penalty function method:}Then, add the transformed rank one constriant to the objective function as a penalty term as shown in Eq. \eqref{Prob:EE_7}.
		\State  \textbf{Problem Solution:}  The tractable form is Standard SDP problem which can be solved through MOSEK optimization toolbox for MATLAB.\\
		\textbf{Output:} Optimal $\mathbf{b}^*$
	\end{algorithmic}
\end{algorithm}
For the optimal beamformers and PAC of NOMA-BF users computed at stage 1, the efficient reflection coefficient of IRS can be obtained through the following  subproblem of optimization 
\begin{subequations}\label{Prob:EE_6}
	\begin{align}
		\mathop {\max }\limits_{\bm{b}} EE =&\mathop  {\max }\limits_{\bm{b}} \sum\limits_{i=1}^I  {\overline{R}_i}-\rho P_{i,T}\\
		\st\ \quad  &  \gamma_{i,k}\geq \gamma^{\min}_{i,k} \label{EE_27-C1},\forall k,i\\
		& |{b}_{n}|= 1, \forall n,\label{ref_IRS5} 
	\end{align}
\end{subequations}
For the efficient solution, it is important to first transform the above formulated problem into tractable form. Let $\bm{\omega}_{i,k}= \mathbf{W}_{i,k}\mathbf{f}_i$ in Eq. \eqref{9}, the received SINR of $r_{i,k}$ can be represented as
 \begin{align}
	&	\gamma_{i,k}=\frac{P_{i} \beta_{i,k}| \mathbf{b}^H\bm{\omega}_{i,k}|^2} {P_i\sum\limits_{l=k+1}^{K}\beta_{i,l}| \mathbf{b}^H\bm{\omega}_{i,k}|^2+\Psi_{i}+\sigma^2}, \label{38}
\end{align}
where, $\Psi_{i}$ will become as follow
\begin{align}
	&\Psi_{i}={\sum\limits_{{\substack{j=1 \\ j\neq i}}}^I|\mathbf{b}^H\bm{\omega}_{j,k}|^2( P_j\sum_{k=1}^{K}\beta_{j,k})}. \label{39}
\end{align}
Then,
\begin{equation}
	\overline{R}^*_i= \sum_{k=1}^{K}BW(\zeta_{i,k}\log_{2}({ \gamma^*_{i,k}})+\Omega_{i,k}), \label{eq:40}
\end{equation}
\begin{align}
	&\overline{R}^*_i= \sum_{k=1}^{K}BW(\zeta_{i,k}\Big[\log_{2}({P_{i} \beta_{i,k}| \mathbf{b}^H\bm{\omega}_{i,k}|^2}) \nonumber\\
	&-\log_{2}({P_i\sum\limits_{l=k+1}^{K}\beta_{i,l}| \mathbf{b}^H\bm{\omega}_{i,k}|^2+\Psi_{i}+\sigma^2})\Big] \nonumber\\
	&+\Omega_{i,k}), \label{eq:41}
\end{align}
Let $\mathbf{B}=\mathbf{b}\mathbf{b}^H$ and  $\bm{W_{i,k}}=\bm{\omega_{i,k}}\bm{\omega_{i,k}^H}$, where $\mathbf{B}\succeq 0$ and rank ($\mathbf{B}$) = 1, $\bm{W_{i,k}}\succeq 0$ and rank ($\bm{W_{i,k}}$) = 1, then $\overline{R}^*_i$ can be expressed as 
\begin{align}
	&\overline{R}^*_i= \sum_{k=1}^{K}BW\Big(\zeta_{i,k}\Big[\log_{2}\big({P_{i} \beta_{i,k} tr(\mathbf{B}\bm{W_{i,k}})}\big) \nonumber\\
	&-\log_{2}\big({P_i\sum\limits_{l=k+1}^{K}\beta_{i,l}tr(\mathbf{B}\bm{W_{i,k}})+\Psi_{i}+\sigma^2}\big)\Big] \nonumber\\
	&+\Omega_{i,k}\Big), \label{eq:42}
\end{align}
Furthermore, $\overline{R}^*_i$ can be presented as a function of $\mathbf{B}$ as
\begin{align}
	&\overline{R}^*_i= \sum_{k=1}^{K}BW\Big(\zeta_{i,k}\Big[f_1(\mathbf{B}) -f_2 (\mathbf{B})\Big]+\Omega_{i,k}\Big), \label{eq:43}
\end{align}
where, 
\begin{align}
	&f_1(\mathbf{B}) =\log_{2}\big({P_{i} \beta_{i,k} tr(\mathbf{B}\bm{W_{i,k}})}\big), \nonumber \\
	&f_2(\mathbf{B}) =\log_{2}\big({P_i\sum\limits_{l=k+1}^{K}\beta_{i,l}tr(\mathbf{B}\bm{W_{i,k}})+\Psi_{i}+\sigma^2}\big) , \label{eq:44}
\end{align}
 It can be observed that $\overline{R}^*_i$ in Eq. \eqref{eq:43} is still not a concave function because it is the difference between two concave functions. Therefore, optimization problem in Eq. \eqref{Prob:EE_6} is not a convex optimization problem. This non-convex optimization problem is transformed into a convex problem by employing low complexity suboptimal technique based on DC programming. In this technique, instead of $f_2(\mathbf{B})$, we substitute its first-order linear approximation as follows
  \begin{align}
  	f_2(\mathbf{B}) & \leq f_2(\mathbf{B}^{(t)})+tr\Big(\big(f_2^{'}(\mathbf{B}^{(t)})\big)^{H}\big(\mathbf{B}-\mathbf{B}^{(t)}\big)\Big), \nonumber\\
  	& \triangleq \overline{f_2(\mathbf{B})} \label{eq:45}
  \end{align} 
where $\mathbf{B}^{(t)}$ is the value of $\mathbf{B}$ at $t^{th}$ iteration and $f_2^{'}(\mathbf{B}^{(t)})$ is the first derivative of $f_2(\mathbf{B})$ at $t^{th}$ iteration, which can be computed as follows
\begin{align}
	f_2^{'}(\mathbf{B}^{(t)}) = \frac {\sum\limits_{l=k+1}^{K}P_i\beta_{i,l}\bm{W_{i,k}}^H}{\Big({P_i\sum\limits_{l=k+1}^{K}\beta_{i,l}tr(\mathbf{B}\bm{W_{i,k}})+\Psi_{i}+\sigma^2}\Big)\ln(2)} \label{eq:46}
\end{align} 
Therefore, $\overline{R}^*_i$ can approximately transformed into the following form
\begin{align}
	&\overline{R}^*_i=  \sum_{k=1}^{K}BW\Big(\zeta_{i,k}\Big[f_1(\mathbf{B}) -\overline{f_2(\mathbf{B})}\Big]+\Omega_{i,k}\Big), \label{eq:47}
\end{align}
 Based on the above computations and approximation, the optimization problem in Eq. \eqref{Prob:EE_6} can be rewritten as
 \begin{subequations}\label{Prob:EE_7}
 	\begin{align}
 		\mathop {\max }\limits_{\bm{b}} EE =&\mathop  {\max }\limits_{\bm{b}} \sum\limits_{i=1}^I  \sum_{k=1}^{K}BW\Big(\zeta_{i,k}\Big[f_1(\mathbf{B}) -\overline{f_2(\mathbf{B})}\Big]+\Omega_{i,k}\Big) \nonumber\\
 		&-\rho P_{i,T}\\
 		\st\ \quad  &  {P_{i} \beta_{i,k} tr(\mathbf{B}\bm{W_{i,k}})}\geq \gamma^{\min}_{i,k} \nonumber \\
 		&\times\big({P_i\sum\limits_{l=k+1}^{K}\beta_{i,l}tr(\mathbf{B}\bm{W_{i,k}})+\Psi_{i}+\sigma^2}\big) \label{EE_6_C1},\forall k,i\\
 		& diag(\mathbf{B}) \leq I_{N},\label{EE_6_C2} \\
 		& \mathbf{B} \succeq 0,\label{EE_6_C3} \\
 		& rank (\mathbf{B}) = 1 \label{EE_6_C4}, 
 	\end{align}
 \end{subequations}
The rank one constraint in the above optimization problem is non-convex which is first transformed in the form of difference of two convex functions as follows
\begin{equation}
	rank (\mathbf{B}) = 1 \Leftrightarrow tr(\mathbf{B}) -||\mathbf{B}||_2 =0,
	\label{eq:49}
\end{equation}
where $tr(\mathbf{B}) = \sum\limits_{n=1}^{N} \lambda_n$ and $\lambda_n$ presents $n^{th}$ largest singular value of $\mathbf{B}$. $||\mathbf{B}||_2$ is the spectral norm of matrix $\mathbf{B}$. The transformed rank one constraint is still non-convex, therefore SCA is exploited to replace $||\mathbf{B}||_2$  with its first order taylor approximation to obtain its lower bound as follows   
\begin{align}
	||\mathbf{B}||_2 &\geq ||\mathbf{B}^{(t)}||_2 tr\Big(\varkappa^{(t)}_{max} {\varkappa_{max}^{(t)}}^H(\mathbf{B}-\mathbf{B}^{(t)})\Big)\nonumber , \\
	&  \triangleq \overline{||\mathbf{B}||_2},\label{eq:50}
\end{align}

where $\varkappa^{(t)}_{max}$ is the eigenvector corresponding to the largest singular value of matrix $\mathbf{B}$  in the $t^{th}$ iteration. Then, add the transformed rank one constraint to the objective function as a penalty term in the optimization problem presented in Eq as follow
\begin{subequations}\label{Prob:EE_7}
	\begin{align}
		\mathop {\max }\limits_{\bm{b}} EE =&\mathop  {\max }\limits_{\bm{b}} \sum\limits_{i=1}^I  \sum_{k=1}^{K}BW\Big(\zeta_{i,k}\Big[f_1(\mathbf{B}) -\overline{f_2(\mathbf{B})}\Big]+\Omega_{i,k}\Big) \nonumber\\
		&-\rho P_{i,T}-\eta(tr(\mathbf{B}) -\overline{||\mathbf{B}||_2})\\
		\st\ &\eqref{EE_6_C1},\eqref{EE_6_C2}, and \eqref{EE_6_C3},   
	\end{align}
\end{subequations}
where, $\eta>>0$ is the penalty factor of the rank one constraint. The above problem is now in the form of standard semi-definite
programming (SDP) problem, which can be solved through the MOSEK optimization toolbox for MATLAB. MOSEK in each iteration solves a relaxed SDP problem through the interior point method, whose computational complexity can be given by $\mathcal{O}(N)^{3.5}$. If $l$ is the number of iterations that are required for the ORCA algorithm to reach convergence. Then, the total computational complexity of the ORCA algorithm is $\mathcal{O}(lN^{3.5})$.
 \section{RESULTS AND DISCUSSION}
 This section presents the simulation results
 of our proposed energy efficient IRS NOMA-BF algorithm for extremely overloaded 6G wireless communications systems. The results are obtained using Monte Carlo simulations. We compare the proposed IRS NOMA-BF approach with IRS conventional BF approach as the benchmark. The considered benchmark refers to multiuser BF systems that serve users orthogonally through BF vectors. In our simulations, the BS has 5 transmit BF vectors, where each BF vector can support a group of two users defined as a cluster. A total of 5 clusters are generated through our proposed clustering from 30 users which are deployed randomly around the IRS through binomial point process (BPP). The minimum distance between BS and IRS is assumed to be 30m while the locations of users around IRS are modelled as binomial point process (BPP) in a circle of radius 10m.For $d_0 = 1 (m)$, $L_0$ is set to -30dB. The path loss exponents are set as 2.2. The number of antennas at BS and the number of elements at IRS are mentioned with each figure. Other main simulation parameters for our setup are described in table I.
  \begin{table}
 	\begin{center}
 		\caption{Simulation Parameters}\label{tbl2}
 		\begin{tabular}{ |c|c|c| }
 			\hline
 			\textbf{Parameter} & \textbf{Value} \\
 			\hline
 			Number of beamforming vectors & 5  \\
 			\hline
 			Number of clusters & 5  \\
 			\hline
 		    Number of users per clusters & 2  \\
 			\hline
 			Users distribution around IRS & BPP  \\
 			\hline
 			IRS radius & 10m  \\
 			\hline
 			path-loss exponent($\alpha$) & 2.2  \\
 			\hline
 			Noise power $(\sigma^2)$ & -114 dBm \\
 			\hline
 			Transmit power of each cluster $(P_i)$ & 30 dBm \\
 			\hline
 			Rician factor $(\delta)$ & 3 dB \\
 			\hline
 			Rician factor $(\epsilon)$ & 3 dB \\
 			\hline
 			The path loss $L_0$ with $d_0 = 1 (m)$ & -30dB\\
 			\hline
 			Distance between IRS and BS $(d_{B,R})$ & 30m \\
 			\hline
 			Circuit power consumption $(P_c)$ & 30 dBm \\
 			\hline
 			minimum SINR $\gamma^{i}_{k,min}$ & 3 dB \\
 			\hline
 			Fast fading & Rician fading\\
 			\hline
 			Pathloss model & distance dependent\\
 			\hline
 			Correlation threshold ($\Delta$) & Between 0 and 1\\
 			\hline
 		\end{tabular}
 	\end{center}
 \end{table}

Figure 3 compares the EE of the proposed IRS NOMA-BF algorithm with IRS conventional BF as the benchmark. The comparison is done with the different number of antennas at BS versus the different number of elements at the IRS. It can be observed from the figure that the proposed algorithm performs better than the benchmark in terms of EE. Moreover, it can be noticed that a higher number of antennas at BS and a higher number of elements at IRS results in higher EE.

 Figure 4 depicts the efficacy of the proposed clustering in terms of ICI reduction while figure 5 presents its efficacy in terms of EE. It can be analyzed from the figures that proposed clustering reduces ICI and thus increases the EE of the IRS NOMA-BF system. From these figures, it can be concluded that it is very important to consider an efficient clustering algorithm among the users which shares the same beamforming vector. Besides it, the higher number of antennas at BS and higher number of elements at IRS increase ICI but as compared to ICI, they greatly improve the desired signal at users. Therefore, the higher number of BS antennas and IRS elements result in higher EE performance.
 
 Figure 6 demonstrates that after efficient clustering the performance of the IRS NOMA-BF system can further be enhanced through efficient PAC of users that share the same BF vector and through efficient allocation of refection coefficients at IRS. The proposed algorithm achieves optimal EE in two stages. Stages 1 is named as optimal power allocation coefficients (OPAC) stage, which allocates efficient PAC to the users. It can be observed from the figure that OPAC attain higher EE than random allocation algorithm. The optimal reflection coefficient algorithm (ORCA) is the second stage of the proposed algorithm which additionally improve the EE of the IRS NOMA-BF system.
 
 Our proposed alternating optimization framework for the IRS NOMA-BF system is consist of two stages, where the first stage employs the OPAC algorithm for optimal PAC of users and 2nd stage also employs an iterative algorithm for optimal RC of IRS. Therefore, it is necessary to analyze the convergence of both OPAC and ORCA algorithms. In Fig. 7, EE convergence of OPAC algorithm versus iterations is demonstrated with the different number of IRS elements (N). The obtained result shows that the OPAC algorithm usually converges in four iterations regardless of N. It is analyzed that N affects EE, but it has a negligible effect on the convergence of the OPAC algorithm. Similarly, in the 2nd stage ORCA algorithm achieve convergence in the $5^{th}$ iteration regardless of the number of IRS elements as shown in Fig. 8.
\begin{figure}
	\centering
	\includegraphics[width=90mm]{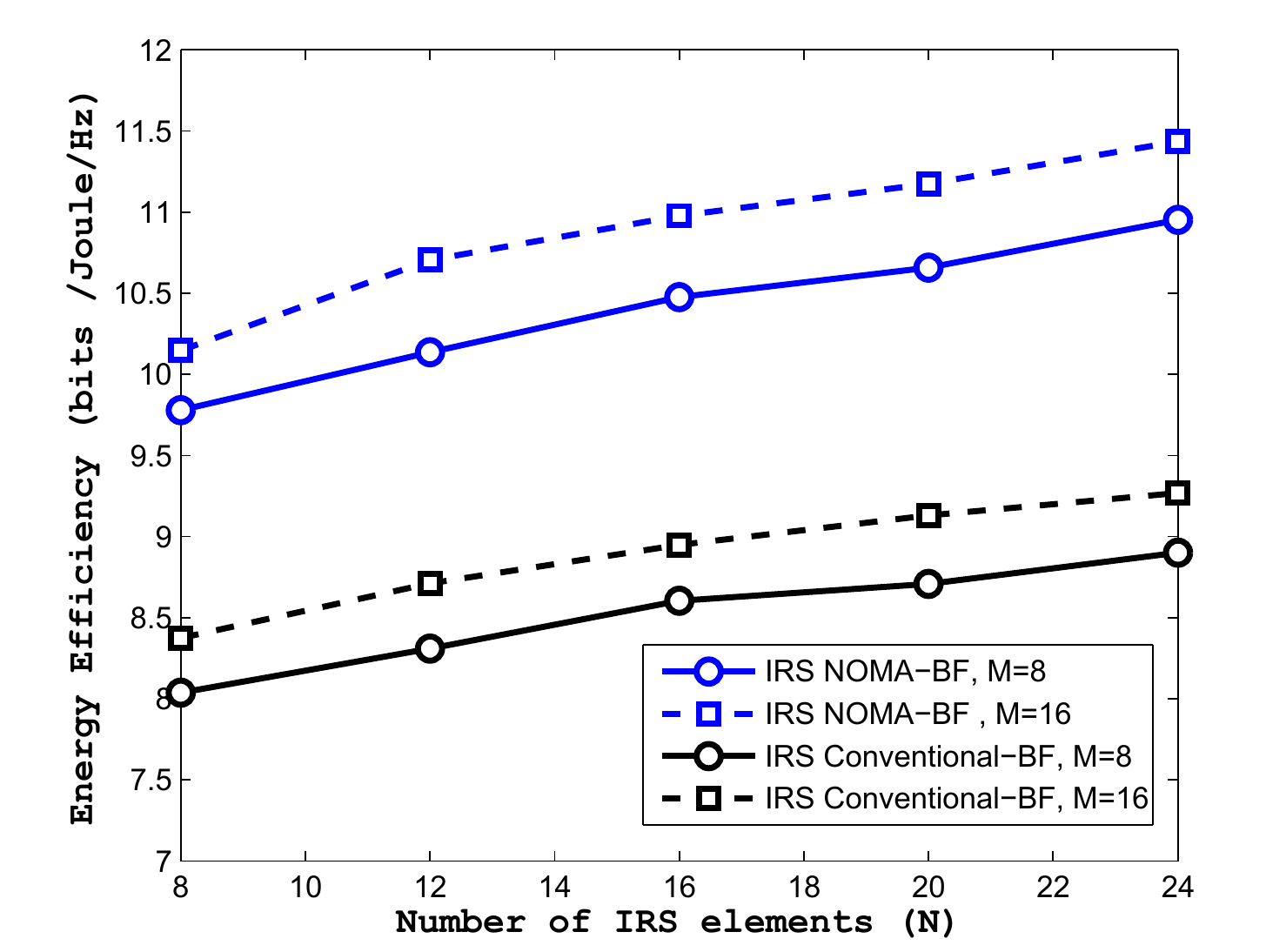}
	\caption{Energy efficiency comparison of proposed algorithm with benchmark algorithm versus number of IRS elements.}
	\label{FIG:3}
\end{figure}
\begin{figure}
	\centering
	\includegraphics[width=90mm]{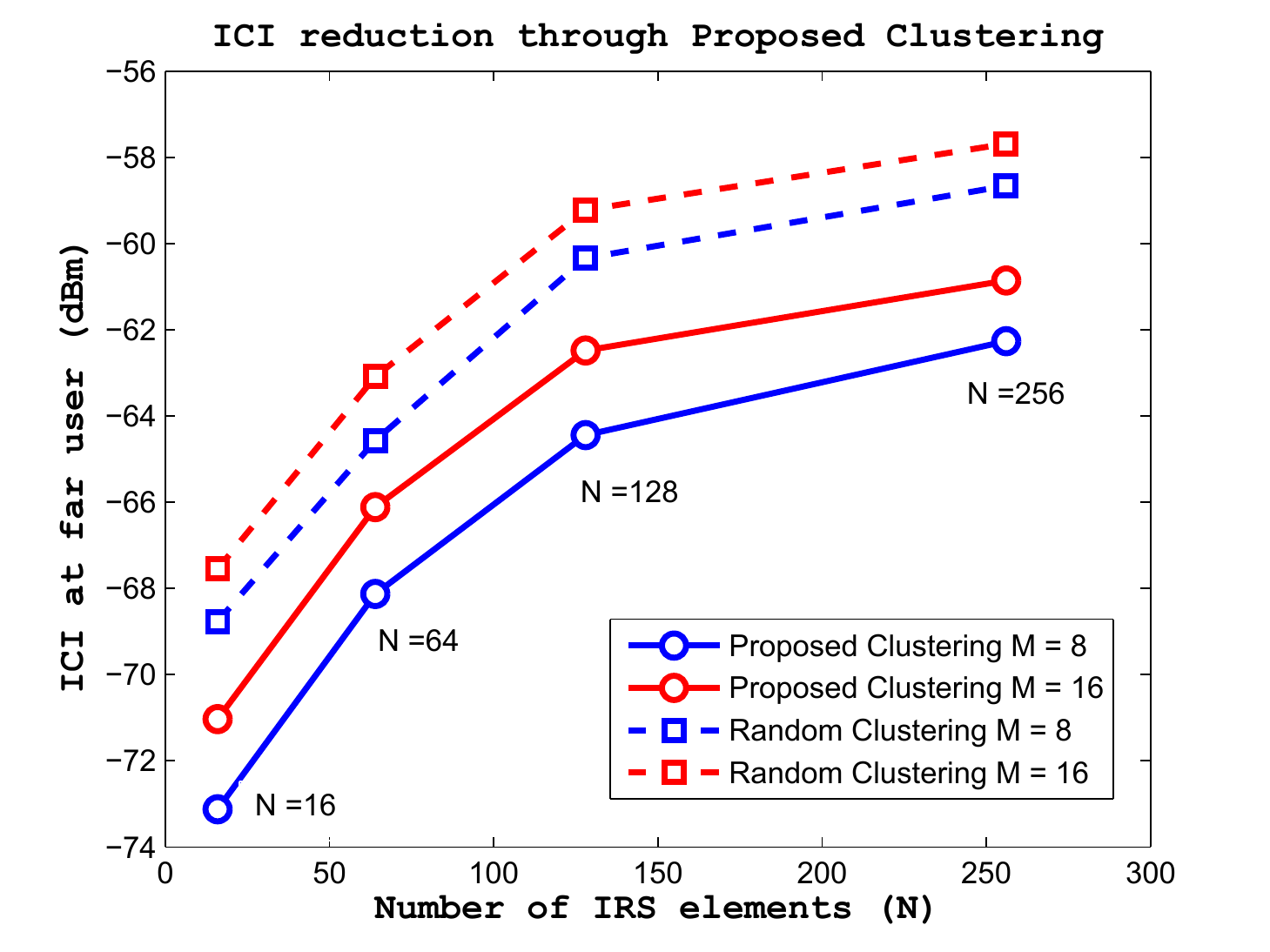}
	\caption{ICI at far user under proposed clustering and random clustering versus number of IRS elements}
	\label{FIG:4}
\end{figure}
\begin{figure}
	\centering
	\includegraphics[width=90mm]{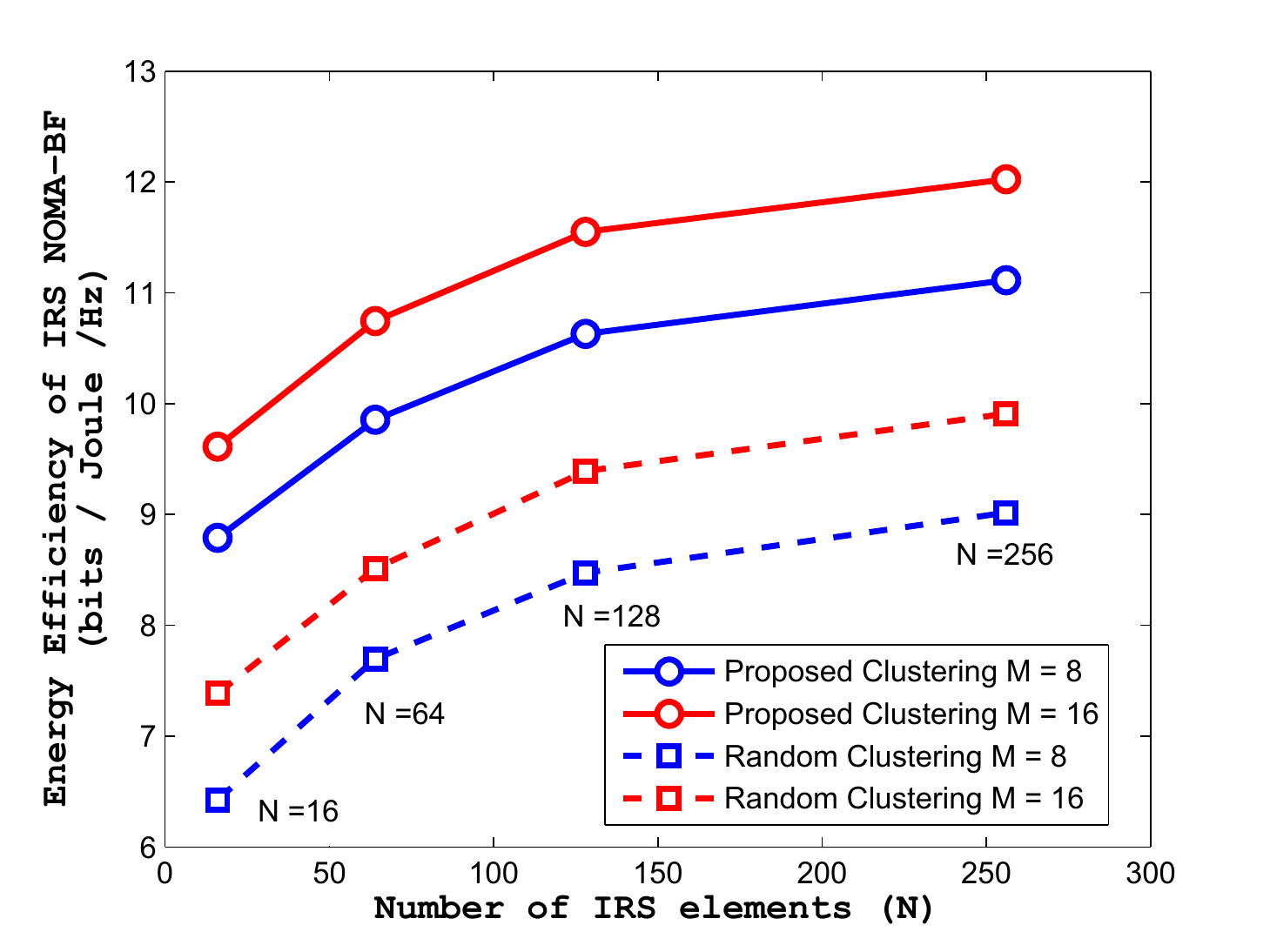}
	\caption{Energy efficiency of IRS NOMA-BF under proposed clustering and random clustering versus number of IRS elements}
	\label{FIG:5}
\end{figure}
\begin{figure}
	\centering
	\includegraphics[width=90mm]{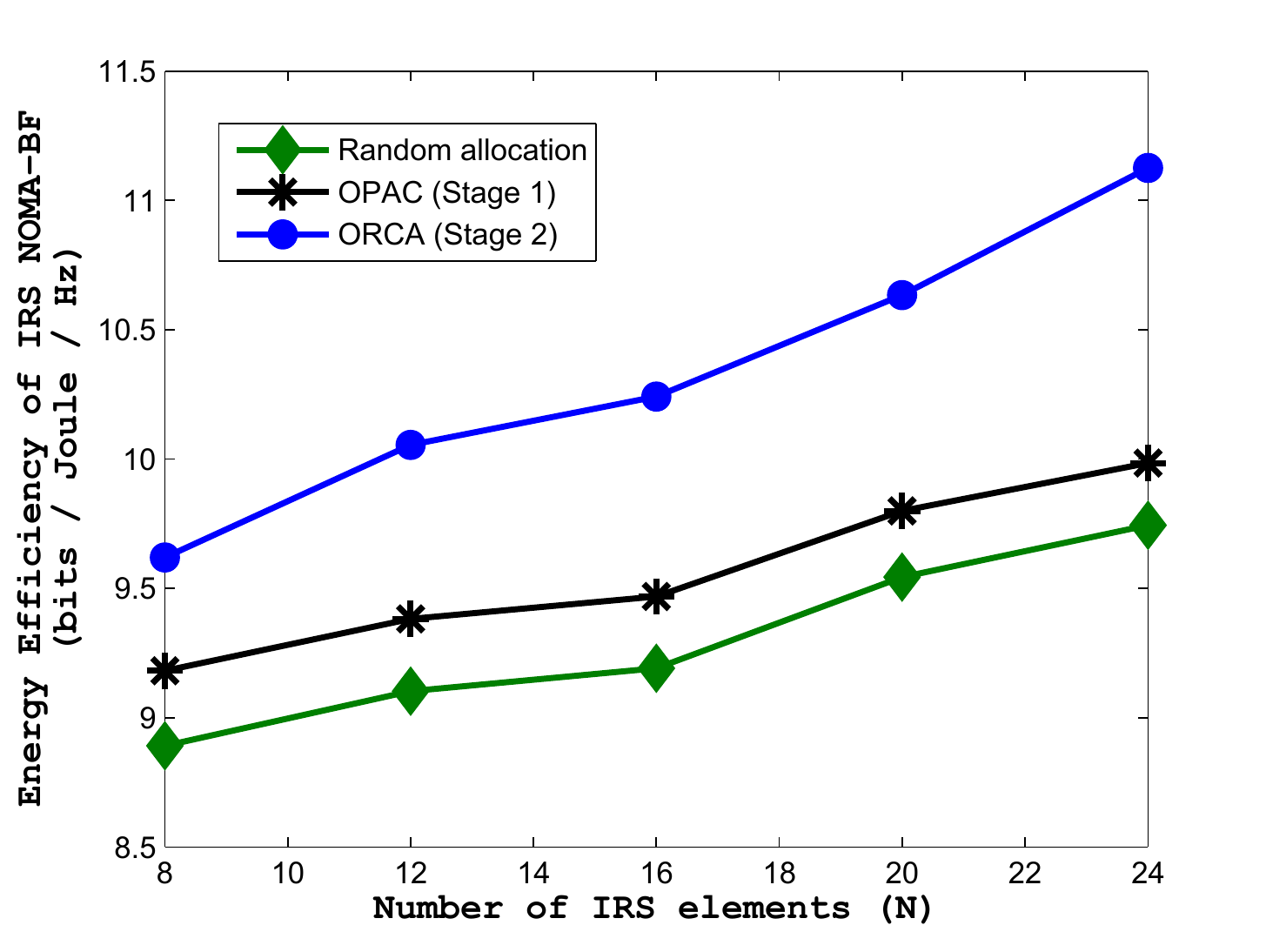}
	\caption{Energy efficiency of proposed algorithm under different stages vesus number of IRS elements}
	\label{FIG:6}
\end{figure} 
\begin{figure}
	\centering
	\includegraphics[width=90mm]{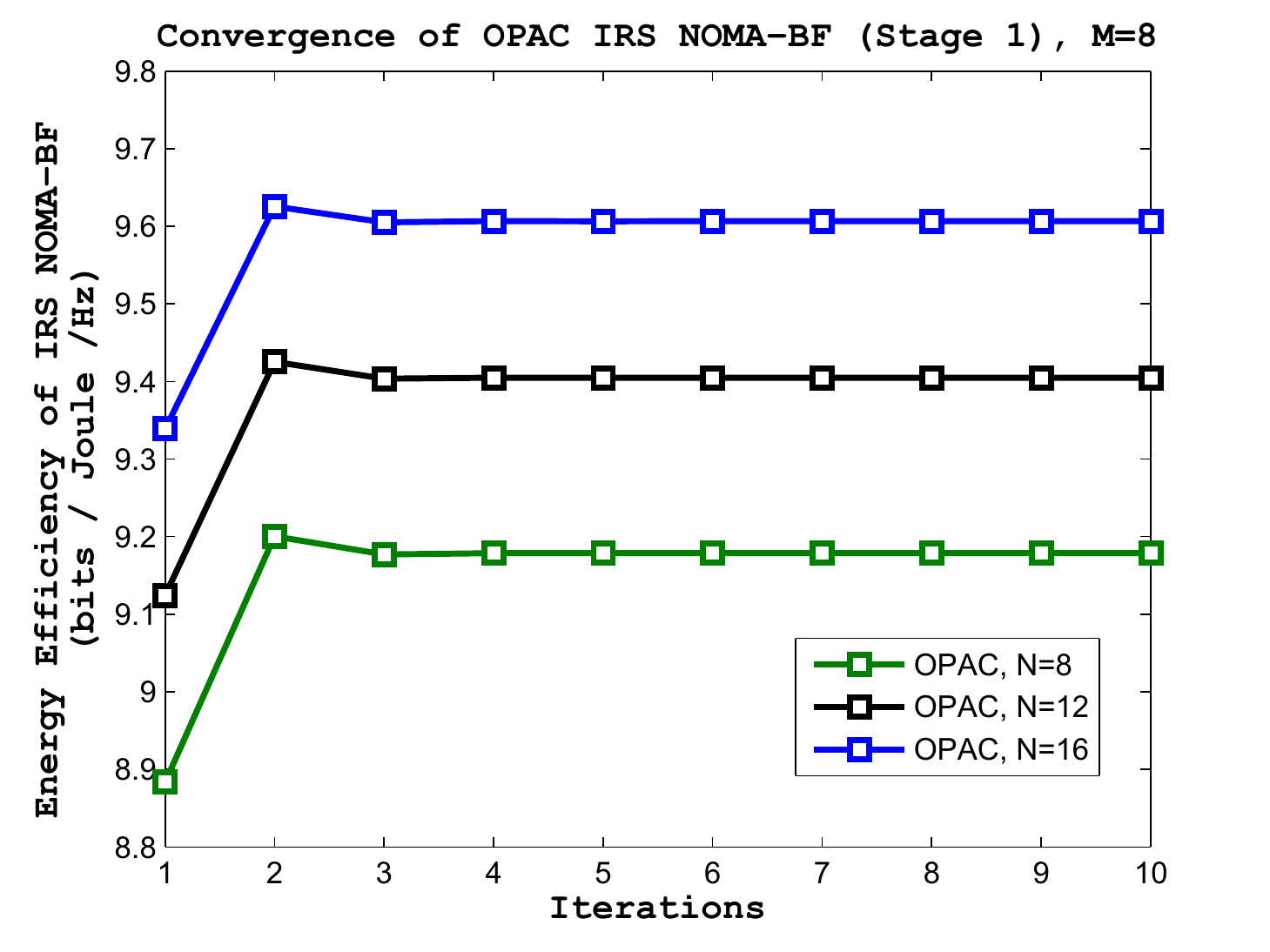}
	\caption{Energy efficiency convergence of OPAC algorithm (Stage 1) for IRS NOMA-BF system}
	\label{FIG:7}
\end{figure}
\begin{figure}
	\centering
	\includegraphics[width=90mm]{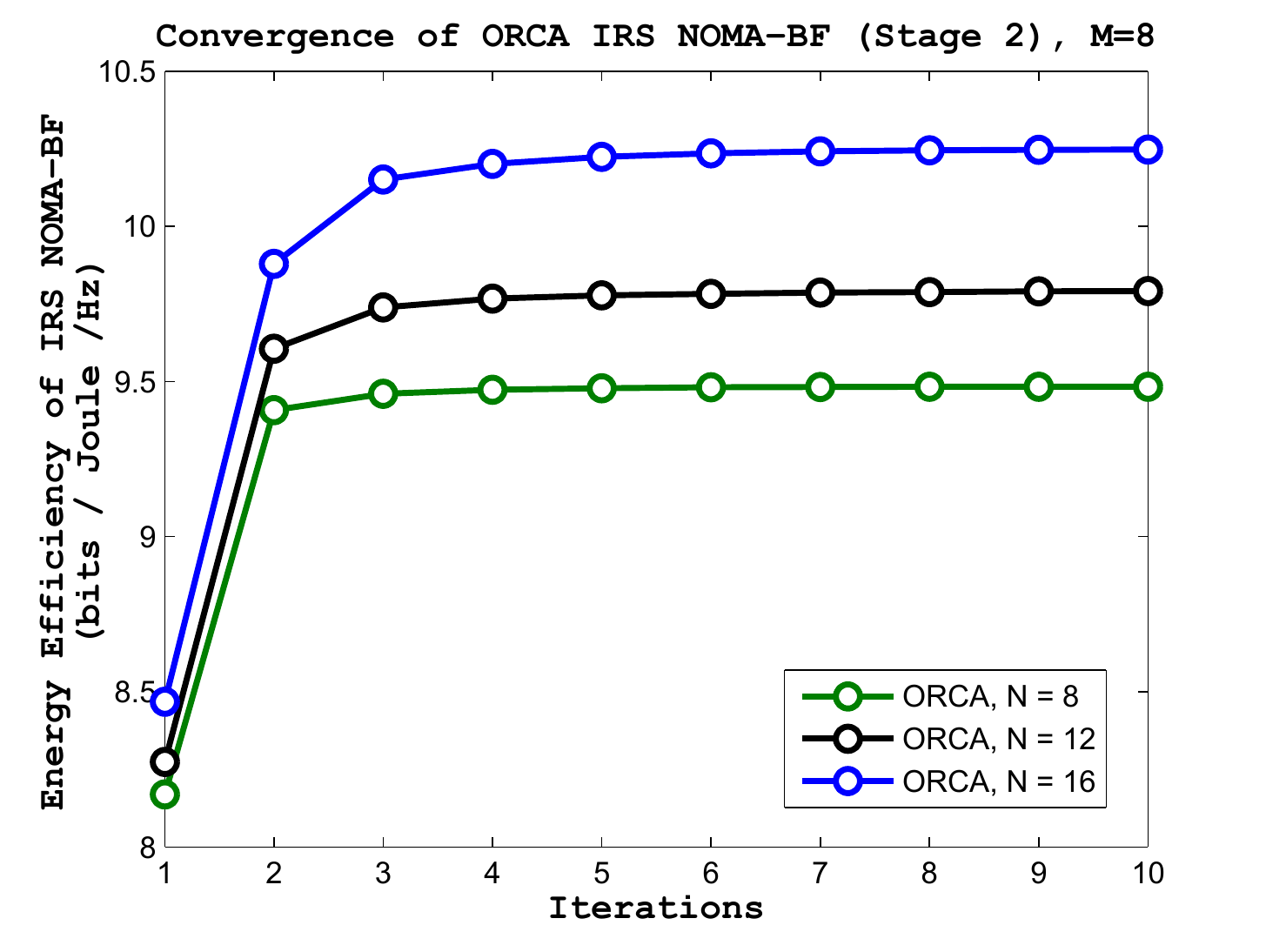}
	\caption{Energy efficiency convergence of ORCA algorithm (Stage 2) for IRS NOMA-BF system}
	\label{FIG:8}
\end{figure}
\section{Conclusion}
A novel alternating optimization framework has been proposed to enhance the energy efficiency of IRS-assisted NOMA-BF systems for next-generation wireless communication networks. More specifically, we aim to maximize the energy efficiency of the considered IRS-assisted NOMA-BF system by optimizing the active beamforming and PAC at the source, and passive beamforming at the IRS node of the system. However, an increment in the number of supportable users with the NOMA-BF system will lead to NOMA user interference and ICI. In this regard, zero-forcing beamforming along with an effective clustering algorithm is exploited to reduce the effect of ICI. Furthermore, a two-stage low-complexity iterative alternating optimization algorithm has been proposed for considered IRS-assisted NOMA-BF system. In the first stage, NOMA user interference is tackled by efficiently optimizing the PAC of NOMA users under the required system constraints. However, in the second stage of the proposed algorithm, passive beamforming has been optimized based on the difference-of-convex (DC) programming and successive convex approximation (SCA).  Moreover, numerical simulation results depict that the proposed alternating optimization framework outperforms its counterpart by providing an efficient performance in terms of the energy efficiency of the system. 
\appendices
\section{Derivation of Optimal PAC of NOMA-BF users at the transmitter}
For efficient SIC process at receivers in each cluster, it is assumed that  users are sorted according to the channel gain as $||\mathbf{u}_{i,k+1}||^2 \geq||\mathbf{u}_{i,k}||^2$  \\
When $k=1$,  the optimal closed-form expression of the first user in the NOMA-BF cluster can be calculated as
\begin{align}
	\frac{\partial\mathcal{L}(\bm{\beta},\alpha, \bm{\varphi},\bm{\Upsilon})}{\partial\beta_{i,1}}= \: &\frac{BW \zeta_{i,k}}{\ln{2}\beta_{i,1}}-\rho_iP_i-\alpha_iP_i+\varphi_{i,1}P_{i}|\mathbf{u}_{i,1}\mathbf{f}_i|^2\nonumber\\
	&+\Upsilon_{i,1}|\mathbf{u}_{i,2}|^2=0,
	\label{eq:52}
\end{align}
Then
\begin{equation}
	\beta_{i,1}=\frac{BW\zeta_{i,1}}{\ln{2}\Big((\rho_i+\alpha_i)P_i-\Gamma_{i,1}\Big)},
	\label{eq:53}
\end{equation}
where
\begin{align}
	\Gamma_{i,1} = \varphi_{i,1}P_{i}|\mathbf{u}_{i,1}\mathbf{f}_i|^2+\Upsilon_{i,1}|\mathbf{u}_{i,2}|^2.  \label{eq:54}
\end{align}
When $k=2$, the optimal closed-form expression of the 2nd user in the NOMA-BF cluster can be computed as

\begin{align}
	&\frac{\partial\mathcal{L}(\bm{\beta},\alpha, \bm{\varphi},\bm{\Upsilon})}{\partial\beta_{i,2}}=\Big(\frac{-BW\zeta_{i,1}P_{i}|\mathbf{u}_{i,1}\mathbf{f}_i|^2}{\ln2\big({P_i|\mathbf{u}_{i,1}\mathbf{f}_i|^2(\beta_{i,2}+\beta_{i,3})+\Psi_{i}+\sigma^2\big)}}\Big)\nonumber\\
	& +\frac{BW\zeta_{i,2}}{\ln2\beta_{i,2}}-\rho_iP_i-\alpha_iP_i-\varphi_{i,1}\gamma^{\min}_{i,1}P_{i}|\mathbf{u}_{i,1}\mathbf{f}_i|^2+\varphi_{i,2}P_{i}|\mathbf{u}_{i,2}\mathbf{f}_i|^2 \nonumber\\
	& +\Upsilon_{i,2}|\mathbf{u}_{i,3}|^2-\Upsilon_{i,1}|\mathbf{u}_{i,2}|^2=0,
	\label{eq:55}
\end{align}
\\
Then we can obtain
\begin{equation}
	\beta_{i,2}=\frac{BW\zeta_{i,2}}{\ln{2}\Big((\rho_i+\alpha_i)P_i-\Gamma_{i,2}\Big)+\Pi(\beta_{i,1})},
	\label{eq:56}
\end{equation}
where
\begin{align}
	\Gamma_{i,2} = \varphi_{i,2}P_{i}|\mathbf{u}_{i,2}\mathbf{f}_i|^2-\Upsilon_{i,2}|\mathbf{u}_{i,3}|^2,  \label{eq:57}
\end{align}
\begin{align}
	\Pi(\beta_{i,1}) =&\frac{BW\zeta_{i,1}}{\ln2\beta_{i,1}}\gamma_{i,1}+\varphi_{i,1}\gamma^{\min}_{i,1}P_{i}|\mathbf{u}_{i,1}\mathbf{f}_i|^2+\Upsilon_{i,1}|\mathbf{u}_{i,2}|^2. \label{eq:58}
\end{align}
\\
When $k=3$, the optimal closed-form expression of the 3rd user in the NOMA-BF cluster can be evaluated as

\begin{align}
	&\frac{\partial\mathcal{L}(\bm{\beta},\alpha, \bm{\varphi},\bm{\Upsilon})}{\partial\beta_{i,3}}=\Big(\frac{-BW\zeta_{i,1}P_{i}|\mathbf{u}_{i,1}\mathbf{f}_i|^2}{\ln2\big({P_i|\mathbf{u}_{i,1}\mathbf{f}_i|^2(\beta_{i,2}+\beta_{i,3})+\Psi_{i}+\sigma^2\big)}}\Big)\nonumber\\
	& \Big(\frac{-BW\zeta_{i,2}P_{i}|\mathbf{u}_{i,2}\mathbf{f}_i|^2}{\ln2\big({P_i|\mathbf{u}_{i,2}\mathbf{f}_i|^2(\beta_{i,3})+\Psi_{i}+\sigma^2\big)}}\Big)+\frac{BW\zeta_{i,3}}{\ln2\beta_{i,3}}-\rho_iP_i-\alpha_iP_i\nonumber\\
	& -\varphi_{i,1}\gamma^{\min}_{i,1}P_{i}|\mathbf{u}_{i,1}\mathbf{f}_i|^2-\varphi_{i,2}\gamma^{\min}_{i,2}P_{i}|\mathbf{u}_{i,2}\mathbf{f}_i|^2+\varphi_{i,3}P_{i}|\mathbf{u}_{i,3}\mathbf{f}_i|^2 \nonumber\\
	& -\Upsilon_{i,2}|\mathbf{u}_{i,3}|^2-\Upsilon_{i,1}|\mathbf{u}_{i,2}|^2=0,
	\label{eq:59}
\end{align}
\\
Then we have

\begin{equation}
	\beta_{i,3}=\frac{BW\zeta_{i,3}}{\ln{2}\Big((\rho_i+\alpha_i)P_i-\Gamma_{i,3}\Big)+\Pi(\beta_{i,1})++\Pi(\beta_{i,2})},
	\label{eq:60}
\end{equation}
where
\begin{align}
	\Gamma_{i,3} = \varphi_{i,3}P_{i}|\mathbf{u}_{i,3}\mathbf{f}_i|^2-\Upsilon_{i,3}|\mathbf{u}_{i,4}|^2,  \label{eq:61}
\end{align}
\begin{align}
	\Pi(\beta_{i,1}) =&\frac{BW\zeta_{i,1}}{\ln2\beta_{i,1}}\gamma_{i,1}+\varphi_{i,1}\gamma^{\min}_{i,1}P_{i}|\mathbf{u}_{i,1}\mathbf{f}_i|^2+\Upsilon_{i,1}|\mathbf{u}_{i,2}|^2, \label{eq:62}
\end{align}
and 
\begin{align}
	\Pi(\beta_{i,2}) =&\frac{BW\zeta_{i,2}}{\ln2\beta_{i,2}}\gamma_{i,2}+\varphi_{i,2}\gamma^{\min}_{i,2}P_{i}|\mathbf{u}_{i,2}\mathbf{f}_i|^2+\Upsilon_{i,2}|\mathbf{u}_{i,3}|^2. \label{eq:63}
\end{align}
Therefore, the closed-form expression for the $k$-th NOMA-BF user in a cluster can be derived as presented in Eq. (\ref{eq:31}).

\bibliographystyle{IEEEtran}

\begin{thebibliography}{10}
\providecommand{\url}[1]{#1}
\csname url@samestyle\endcsname
\providecommand{\newblock}{\relax}
\providecommand{\bibinfo}[2]{#2}
\providecommand{\BIBentrySTDinterwordspacing}{\spaceskip=0pt\relax}
\providecommand{\BIBentryALTinterwordstretchfactor}{4}
\providecommand{\BIBentryALTinterwordspacing}{\spaceskip=\fontdimen2\font plus
\BIBentryALTinterwordstretchfactor\fontdimen3\font minus
  \fontdimen4\font\relax}
\providecommand{\BIBforeignlanguage}[2]{{%
\expandafter\ifx\csname l@#1\endcsname\relax
\typeout{** WARNING: IEEEtran.bst: No hyphenation pattern has been}%
\typeout{** loaded for the language `#1'. Using the pattern for}%
\typeout{** the default language instead.}%
\else
\language=\csname l@#1\endcsname
\fi
#2}}
\providecommand{\BIBdecl}{\relax}
\BIBdecl

\bibitem{R. Alghamdi}
R. Alghamdi et al., "Intelligent Surfaces for 6G Wireless Networks: A Survey of Optimization and Performance Analysis Techniques," in IEEE Access, vol. 8, pp. 202795-202818, 2020.

\bibitem{Q. Wu 1}
Q. Wu and R. Zhang, "Towards Smart and Reconfigurable Environment: Intelligent Reflecting Surface Aided Wireless Network," in IEEE Communications Magazine, vol. 58, no. 1, pp. 106-112, January 2020.

\bibitem{C. Huang}
C. Huang et al., "Holographic MIMO Surfaces for 6G Wireless Networks: Opportunities, Challenges, and Trends," in IEEE Wireless Communications, vol. 27, no. 5, pp. 118-125, October 2020.

\bibitem{Z. Li}
Z. Li, Wen Chen, and H. Cao, “Beamforming Design and Power Allocation for Transmissive RMS-based Transmitter Architectures,” IEEE Wireless Communications Letters, vol. 11, no. 1, pp. 53-57, 2022

\bibitem{Q. Wu 2}
Q. Wu and R. Zhang, “Intelligent reflecting surface enhanced wireless network via joint active and passive beamforming,”
IEEE Trans. Wireless Commun., vol. 18, no. 11, pp. 5394–5409, Nov. 2019.

\bibitem{D. Xu}
 D. Xu, X. Yu, Y. Sun, D. W. K. Ng, and R. Schober, “Resource allocation for IRS-assisted full-duplex cognitive radio
systems,” IEEE Trans. Commun., vol. 68, no. 12, pp. 7376–7394, Dec. 2020.

\bibitem{X. Yu}
X. Yu, D. Xu, Y. Sun, D. W. K. Ng, and R. Schober, “Robust and secure wireless communications via intelligent reflecting
surfaces,” IEEE J. Sel. Areas Commun., vol. 38, no. 11, pp. 2637–2652, Nov. 2020.

\bibitem{Liu}
Liu, Yuanwei, Shuowen Zhang, Xidong Mu, Zhiguo Ding, Robert Schober, Naofal Al-Dhahir, Ekram Hossain, and Xuemin Shen. "Evolution of NOMA toward next generation multiple access (NGMA)." arXiv preprint arXiv:2108.04561 (2021).

\bibitem{A. Ihsan}
A. Ihsan, W. Chen, S. Zhang and S. Xu, "Energy-Efficient NOMA Multicasting System for Beyond 5G Cellular V2X Communications With Imperfect CSI," in IEEE Transactions on Intelligent Transportation Systems

\bibitem{Z. Ding}
Z. Ding and H. Vincent Poor, "A Simple Design of IRS-NOMA Transmission," in IEEE Communications Letters, vol. 24, no. 5, pp. 1119-1123, May 2020.

\bibitem{M. Fu}
M. Fu, Y. Zhou, and Y. Shi, “Intelligent reflecting surface for downlink non-orthogonal multiple access networks,” in Proc.
IEEE Globecom Workshops, Waikoloa, HI, Dec. 2019, pp. 1–6

\bibitem{X. Mu}
X. Mu, Y. Liu, L. Guo, J. Lin and N. Al-Dhahir, "Exploiting Intelligent Reflecting Surfaces in NOMA Networks: Joint Beamforming Optimization," in IEEE Transactions on Wireless Communications, vol. 19, no. 10, pp. 6884-6898, Oct. 2020.

\bibitem{Y. Li}
Y. Li, M. Jiang, Q. Zhang and J. Qin, "Joint Beamforming Design in Multi-Cluster MISO NOMA Reconfigurable Intelligent Surface-Aided Downlink Communication Networks," in IEEE Transactions on Communications, vol. 69, no. 1, pp. 664-674, Jan. 2021.

\bibitem{B. Zheng}
B. Zheng, Q. Wu, and R. Zhang, “Intelligent reflecting surface-assisted multiple access with user pairing: NOMA or
OMA?” IEEE Commun. Lett., vol. 24, no. 4, pp. 753–757, Jan. 2020.

\bibitem{X. Mu 2}
X. Mu, Y. Liu, L. Guo, J. Lin and H. V. Poor, "Intelligent Reflecting Surface Enhanced Multi-UAV NOMA Networks," in IEEE Journal on Selected Areas in Communications, vol. 39, no. 10, pp. 3051-3066, Oct. 2021.

\bibitem{Z. Li 2}
Z. Li, W. Chen, Q. Wu, K. Wang and J. Li, "Joint Beamforming Design and Power Splitting Optimization in IRS-Assisted SWIPT NOMA Networks," in IEEE Transactions on Wireless Communications,2022.

\bibitem{Q. Wu 3}
Q. Wu, X. Zhou and R. Schober, "IRS-Assisted Wireless Powered NOMA: Do We Really Need Different Phase Shifts in DL and UL?," in IEEE Wireless Communications Letters, vol. 10, no. 7, pp. 1493-1497, July 2021.

\bibitem{Q. Wang}
Q. Wang, F. Zhou, H. Hu and R. Q. Hu, "Energy-Efficient Design for IRS-Assisted MEC Networks with NOMA," 2021 13th International Conference on Wireless Communications and Signal Processing (WCSP), pp. 1-6, 2021.

\bibitem{J. Zuo1}
J. Zuo, Y. Liu, L. Yang, L. Song and Y. -C. Liang, "Reconfigurable Intelligent Surface Enhanced NOMA Assisted Backscatter Communication System," in IEEE Transactions on Vehicular Technology, vol. 70, no. 7, pp. 7261-7266, July 2021.

\bibitem{C. Gong}
C. Gong et al., "Intelligent Reflecting Surface Aided Secure Communications for NOMA Networks," in IEEE Transactions on Vehicular Technology, 2022.

\bibitem{M. Z. Chowdhury}
M. Z. Chowdhury, M. Shahjalal, S. Ahmed and Y. M. Jang, "6G Wireless Communication Systems: Applications, Requirements, Technologies, Challenges, and Research Directions," in IEEE Open Journal of the Communications Society, vol. 1, pp. 957-975, 2020.

\bibitem{B. Kimy}
B. Kimy et al., "Non-orthogonal Multiple Access in a Downlink Multiuser Beamforming System," MILCOM 2013 - 2013 IEEE Military Communications Conference, pp. 1278-1283, 2013.

\bibitem{Liu Yuanwei}
Liu, Yuanwei, et al. "Reconfigurable intelligent surface (RIS) aided multi-user networks: Interplay between NOMA and RIS." arXiv preprint arXiv:2011.13336 (2020).

\bibitem{J. Zuo}
J. Zuo, Y. Liu, E. Basar and O. A. Dobre, "Intelligent Reflecting Surface Enhanced Millimeter-Wave NOMA Systems," in IEEE Communications Letters, vol. 24, no. 11, pp. 2632-2636, Nov. 2020.

\bibitem{M. S. Ali}
M. S. Ali, H. Tabassum and E. Hossain, "Dynamic User Clustering and Power Allocation for Uplink and Downlink Non-Orthogonal Multiple Access (NOMA) Systems," in IEEE Access, vol. 4, pp. 6325-6343, 2016.

\bibitem{Q. Shi}
Q. Shi, L. Liu, W. Xu and R. Zhang, "Joint Transmit Beamforming and Receive Power Splitting for MISO SWIPT Systems," in IEEE Transactions on Wireless Communications, vol. 13, no. 6, pp. 3269-3280, June 2014.

\bibitem{J. Papandriopoulos}
J. Papandriopoulos and J. S. Evans, “SCALE: A low-complexity
distributed protocol for spectrum balancing in multiuser DSL networks,” IEEE Trans. Inf. Theory, vol. 55, no. 8, pp. 3711–3724, Aug. 2009.

\bibitem{K. Shen}
K. Shen and W. Yu, “Fractional programming for communication
systems—Part I: Power control and beamforming,” IEEE Trans. Signal Process., vol. 66, no. 10, pp. 2616–2630, May 2018.

\bibitem{M. R. Zamani}
 M. R. Zamani, M. Eslami, M. Khorramizadeh, and Z. Ding, “Energyefficient power allocation for NOMA with imperfect CSI,” IEEE Trans.Veh. Technol., vol. 68, no. 1, pp. 1009–1013, Jan. 2019.


\end{thebibliography}

\begin{IEEEbiography}[{\includegraphics[width=1in,height=1.25in,clip,keepaspectratio]{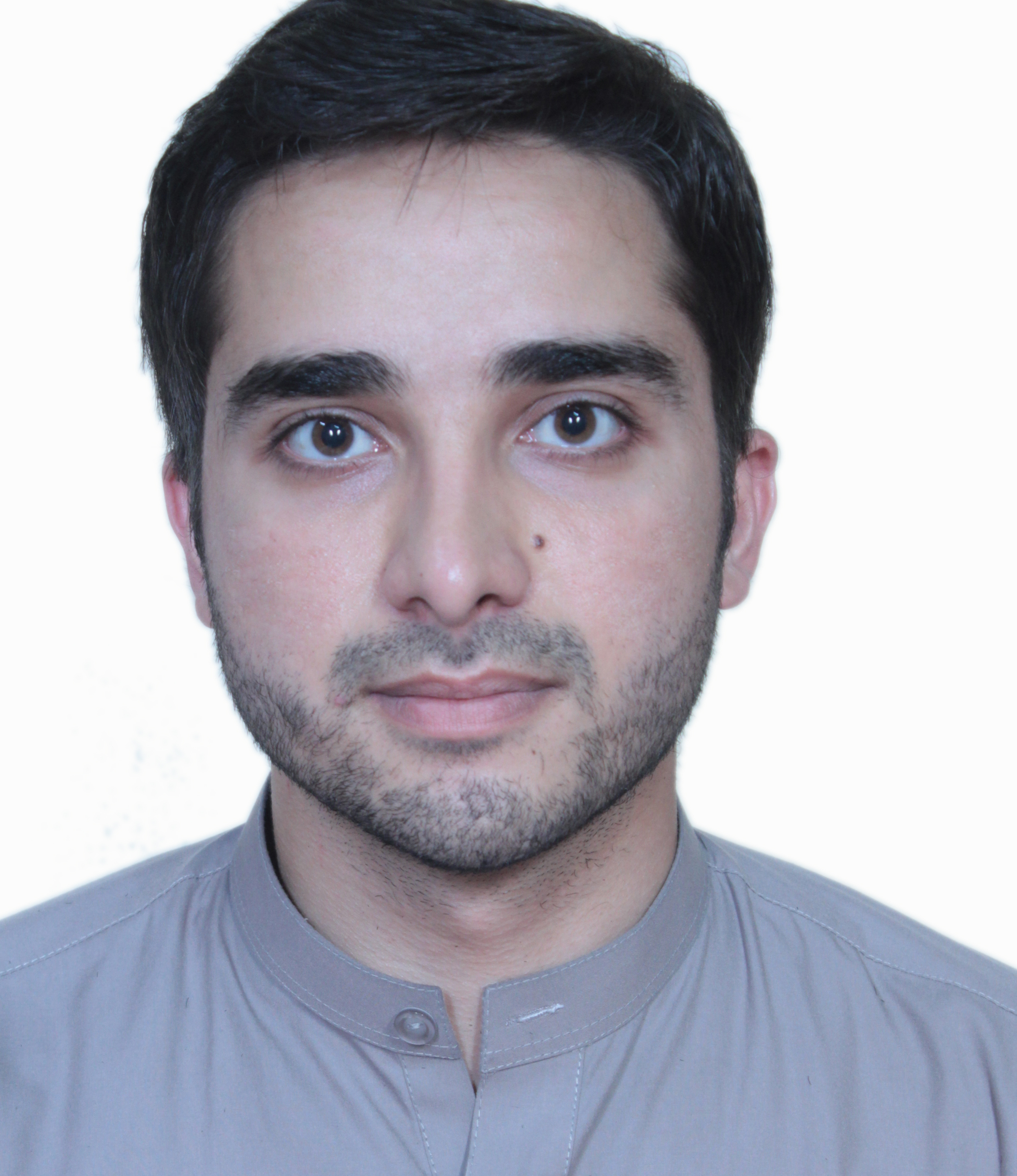}}]{Asim Ihsan}
	received the B.S. degree in Telecommunication Engineering from the University of Engineering and Technology (UET), Peshawar, Pakistan, in 2015, and the M.S. degree in Information and Communication Engineering from Xi’an Jiaotong University (XJTU), Xi’an, China, in 2018. He is
	currently pursuing the Ph.D. degree in Information and Communication Engineering with Shanghai Jiao Tong University (SJTU), Shanghai, China. His current research interests include the intelligent reflective surfaces, internet of vehicles, backscatter communications, physical layer security, and wireless sensor networks. He is an active reviewer of peer reviewed international journals.
\end{IEEEbiography}
\begin{IEEEbiography}[{\includegraphics[width=1in,height=1.25in,clip,keepaspectratio]{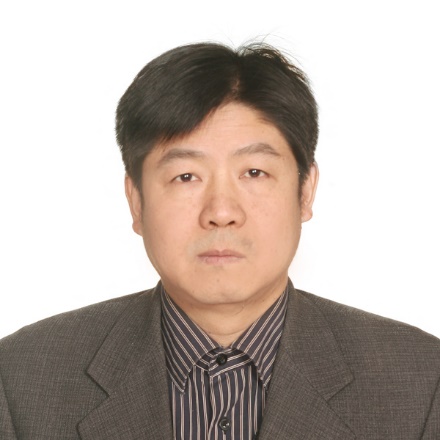}}]{Wen Chen}
	(Senior Member, IEEE) is a tenured Professor with the Department of Electronic Engineering, Shanghai Jiao Tong University, China, where he is the director of Broadband Access Network Laboratory. He is a fellow of Chinese Institute of Electronics and the distinguished lecturers of IEEE Communications Society and IEEE Vehicular Technology Society. He is the Shanghai Chapter Chair of IEEE Vehicular Technology Society, an Editors of IEEE Transactions on Wireless Communications, IEEE Transactions on Communications, IEEE Access and IEEE Open Journal of Vehicular Technology. His research interests include multiple access, wireless AI and meta-surface communications. He has published more than 100 papers in IEEE journals and more than 100 papers in IEEE Conferences, with citations more than 6000 in google scholar.
\end{IEEEbiography}
\begin{IEEEbiography}
	[{\includegraphics[width=1in,height=1.5in,clip,keepaspectratio]{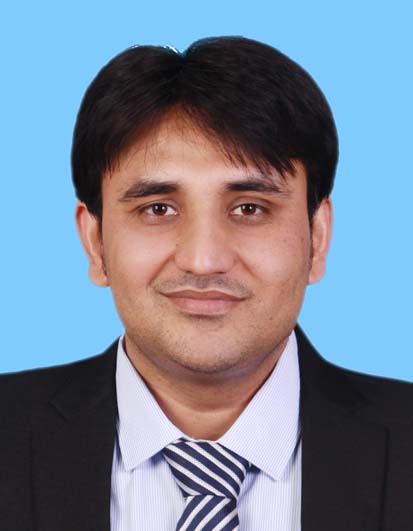}}]{MUHAMMAD ASIF} 
	was born in Rahim Yar Khan, Bahawalpur Division, Pakistan, in 1990. He received the Bachelor of Science (B.Sc) degree in Telecommunication Engineering from The Islamia University of Bahawalpur (IUB), Punjab, Pakistan, in 2013, and Master degree in Communication and Information Systems from Northwestern Polytechnical University (NWPU), Xian, Shaanxi, China, in 2015. He also received Ph.D. degree in Information and Communication Engineering from University of Science and Technology of China (USTC), Hefei, Anhui, China in 2019. Currently, Dr. Asif is working as a post-doctoral researcher at the Department of Electronics and Information Engineering in Shenzhen University, Shenzhen, Guangdong, China. He has authored/co-authored several journal and conference papers. His research interests include Wireless Communication, Channel Coding, Coded-Cooperative Communication, Optimization and Resource Allocation, Backscatter-Enabled Wireless Communication, IRS-Assisted Next-generation IOT Networks.
\end{IEEEbiography}
\begin{IEEEbiography}
	[{\includegraphics[width=1in,height=1.5in,clip,keepaspectratio]{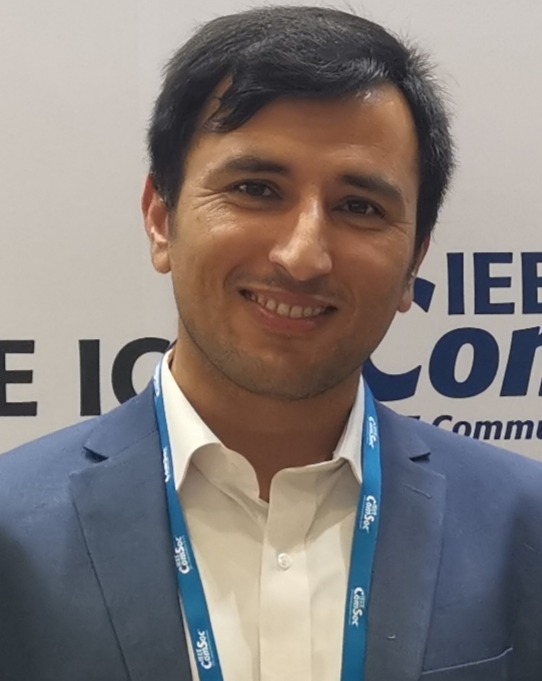}}]{Wali Ullah Khan} (Member, IEEE)
	received the Master degree in Electrical Engineering from COMSATS University Islamabad, Pakistan, in 2017, and the Ph.D. degree in Information and Communication Engineering from Shandong University, Qingdao, China, in 2020. He is currently working with the Interdisciplinary Centre for Security, Reliability and Trust (SnT), University of Luxembourg, Luxembourg. He has authored/coauthored more than 50 publications, including international journals, peer-reviewed conferences, and book chapters. His research interests include convex/nonconvex optimizations, non-orthogonal multiple access, reflecting intelligent surfaces, ambient backscatter communications, Internet of things, intelligent transportation systems, satellite communications, physical layer security, and applications of machine learning.
\end{IEEEbiography}
\vskip -2\baselineskip plus -1fil
\begin{IEEEbiography}
	[{\includegraphics[width=1in,height=1.5in,clip,keepaspectratio]{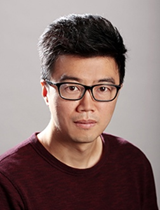}}]{Jun Li} (Senior Member, IEEE)
	received the Ph.D. degree in electronic engineering from Shanghai JiaoTong University, Shanghai, China, in 2009. From January 2009 to June 2009, he worked with the Department of Research and Innovation, Alcatel Lucent Shanghai Bell as a Research Scientist. From June 2009 to April 2012, he was a Post-Doctoral Fellow with the School of Electrical Engineering and Telecommunications, The University of New South Wales, Sydney, NSW, Australia. From April 2012 toJune 2015, he was a Research Fellow with the School of Electrical Engineering, The University of Sydney, Sydney, NSW, Australia. Since June 2015, he has been a Professor with the School of Electronic and Optical Engineering, Nanjing University of Science and Technology, Nanjing, China. He was a Visiting Professor with Princeton University from 2018 to 2019. He has coauthored more than 200 papers in IEEE journals and conferences, and holds one U.S. patents and more than ten Chinese patents in these areas. His research interests include network information theory, game theory, distributed intelligence, multiple agent reinforcement learning, and their applications in ultra-dense wireless networks, mobile edge computing, network privacy and security, and the Industrial Internet of Things. He was serving as an Editor of IEEE COMMUNICATION LETTERS and a TPC member for several flagship IEEE conferences. He received Exemplary Reviewer of IEEE TRANSACTIONS ON COMMUNICATIONS in 2018 and the Best Paper Award from IEEE International Conference on 5G for Future Wireless Networks in 2017.
\end{IEEEbiography}

\end{document}